\documentclass[a4paper,11pt]{article}
\pdfoutput=1 
\usepackage{jcappub}
\usepackage[T1]{fontenc}
\author[a]{Asta~Heinesen} 

\affiliation[a]{Univ Lyon, Ens de Lyon, Univ Lyon1, CNRS, Centre de Recherche Astrophysique de Lyon UMR5574, F--69007, Lyon, France}

\emailAdd{asta.heinesen@ens--lyon.fr}

\usepackage{braket}
\usepackage{changes}
\usepackage{subcaption}
\usepackage[section]{placeins}
\usepackage[titletoc]{appendix}
\usepackage{xcolor}
\usepackage{amssymb}
\usepackage{dsfont}
\usepackage{bm}
\usepackage{mathtools}
\usepackage{enumerate}
\DeclareMathAlphabet{\mathpzc}{OT1}{pzc}{m}{it}

\let\mathbb\mathds

\newcommand{\dd}{{\mathrm d}}
\def\rad{\mathpzc{r}}

\newcommand{\say}[1]{`#1'}
\DeclarePairedDelimiter\abs{\lvert}{\rvert}%

\providecommand{\href}[2]{#2}\def\link#1{\href{http://arxiv.org/abs/#1}{{\tt #1}}}
\def\etal{{et al}.}

\def\lsim{\mathop{\hbox{${\lower3.8pt\hbox{$<$}}\atop{\raise0.2pt\hbox{$\sim$}}
$}}}
\definecolor{MyB}{rgb}{0.1,0.1,1.0}

\title{Cosmological homogeneity scale estimates are dressed}

\abstract{
We investigate number count statistics as measures for transition to homogeneity of the matter distribution in the Universe and analyse how such statistics might be \say{dressed} by the assumed survey selection function.
Since the estimated survey selection function---which ideally accounts for selection bias in the observed distribution---is {partially} degenerate with the estimated underlying distribution of galaxies, the ability to identify the correct survey selection function is of importance for obtaining reliable estimates for clustering statistics. 
Selection functions of existing galaxy catalogues are modelled from data to resemble the redshift distribution and mean density of the observed galaxies. Proposed estimates of the selection function for upcoming surveys in addition {use the angular distribution of galaxies to generate the angular selection function instead of using angular completeness estimates}. We argue that such modelling of the selection function could potentially underestimate the deviance from homogeneity at scales probed by existing catalogues.  

We investigate the impact of conventionally applied methods for estimation of the survey selection function on number count in sphere statistics in a toy model setting. 
The example density distribution is asymptotically homogeneous, while non-linear density fluctuations are present regionally. 
We find that density oscillations with period comparable to characteristic scales of the survey are suppressed when conventional estimates of the survey selection function are invoked, resulting in number count statistics that are biased towards homogeneity. 
For our concrete toy model with maximum density contrasts of $1$ and period of the density oscillation comparable in size to the survey radius, we find that the homogeneity scale is underestimated by $\sim 40\%$, however this quantitative result is dependent on the model setup including the form of the simplistic density field considered and the survey geometry.

}
\catcode`\@=11
\catcode`\@=12
\begin{document}
\maketitle

\def\PRL#1{{\em Phys.\ Rev.\ Lett.}\ {\bf#1}}
\def\JCAP#1{{\em J.\ Cosmol.\ Astropart.\ Phys.}\ {\bf#1}}
\def\ApJ#1{{\em Astrophys.\ J.}\ {\bf#1}}
\def\PR#1{{\em Phys.\ Rev.}\ {\bf#1}}
\def\MNRAS#1{{\em Mon.\ Not.\ R.\ Astr.\ Soc.}\ {\bf#1}}
\def\CQG#1{{\em Class.\ Quantum Grav.}\ {\bf#1}}
\def\GRG#1{{\em Gen.\ Relativ.\ Grav.}\ {\bf#1}}
\def\IJMP#1{{\em Int.\ J.\ Mod.\ Phys.}\ {\bf#1}}
\def\AaA#1{{\em Astron.\ Astrophys}.\ {\bf#1}}
\def\AJ#1{{\em Astron.\ J}.\ {\bf#1}}
\def\ApJs#1{{\em Astrophys.\ J.\ Suppl}.\ {\bf#1}}
\def\PLB#1{{\em Phys.\ Lett.}\ {\bf B #1}}
\def\APSS#1{{\em Astrophys. \ Space \ Sci.}\ {\bf#1}}
\def\NatA#1{{\em Nature Astron.}\ {\bf#1}}
\def\ARNPS#1{{\em Ann.\ Rev.\ Nucl.\ Part.\ Sci.}\ {\bf#1}}

\def\beq{\begin{equation}} \def\eeq{\end{equation}}
\def\bea{\begin{eqnarray}} \def\eea{\end{eqnarray}}
\def\e{\mathop{\rm e}\nolimits}
\def \domain{\mathcal{D}}
\def\doubleunderline#1{\underline{\underline{#1}}}

\DeclareRobustCommand{\orderof}{\ensuremath{\mathcal{O}}}

\tableofcontents
\newpage

\section{Introduction}
Number count statistics are important probes of large scale structure in cosmology, and are for instance used for examining baryon acoustic oscillation {(BAO)} features in cosmological datasets \cite{EisensteinDetection,Cole} and for investigating transition to homogeneity \cite{Hogg,Scrimgeour,Laurent}. 

Large scale cosmological surveys do not perfectly sample galaxies and other astrophysical objects, but are subject to selection effects specific to the instrumentation, foreground contamination, and the luminosity distribution of the observed objects \cite{BlakeBrough,Reid}. 
Selection on the basis of colour and magnitude of galaxies can partly account for non-uniformity in sampling \cite{Eisensteintarget}, but inhomogeneities in sampling in modern large scale surveys remain and must ideally be modelled to avoid systematic errors in estimates derived from the catalogues. 

In modern analysis of number count statistics normalisation by random catalogue number counts are invoked, where the random catalogues are generated from an estimated survey selection function that accounts for radial and angular selection effects \cite{Zehavi,BlakeBrough}.  
Because of the difficulty in modelling the true underlying selection function of objects such as galaxies, it is conventional to estimate the model selection function in redshift directly from the observed density-distribution, which is for instance done for the SDSS-III Baryon Oscillation Spectroscopic Survey (BOSS) of galaxies \cite{Reid} and quasars \cite{Ata} respectively. 
The angular selection can be determined from the estimated observational completeness within angular patches of the sky, as is the case for e.g. the SDSS-III BOSS survey \cite{Reid}. For upcoming surveys from the Dark Energy Spectroscopic Instrument (DESI) experiment \cite{DESI}, it has been proposed to model the angular selection function to resemble the angular distribution of the survey density field \cite{Burden}. 

Normalisation of the \say{bare} galaxy number counts\footnote{In this paper we use the term 'bare' galaxy number counts for number counts which are computed directly from the survey. 'Dressed' galaxy number counts refer to the situation where the number counts have been corrected in accordance with an assumed survey selection function.} has the potential of removing selection effects. However, if the survey selection function is incorrectly estimated, the normalisation operation can \emph{introduce} systematic errors into number count statistics.  
It is a potential concern that an estimated selection function empirically determined from data might account for \emph{physical} large scale inhomogeneity present in the matter distribution, and that number counts that are corrected based on such an estimated selection function might be systematically biased towards a higher degree of homogeneity and isotropy than what is present in the physical Universe. 
It is the aim of this paper to examine the impact of typical estimation methods for model selection functions on number count statistics and homogeneity scale measures. 

The most common number count estimator for transition to homogeneity is the average number counts in spheres (or the analogous \say{correlation dimension}) as employed in, e.g., \cite{Hogg,Scrimgeour,Laurent,Goncalves,Ntelis} (see also \cite{Avila1,Avila2} for studies of angular projections of the estimator). 
The average number count in spheres is the lowest order measure of inhomogeneity and describes the 2-point self-correlation of the density field, and must be supplemented with higher order correlation functions \cite{PeeblesNpoint} in order to fully quantify the density distribution and its transition to homogeneity.  
Supplementary clustering statistics include Minkowski functionals \cite{MeckeBuchertWagner}, which have been used to demonstrate that higher-order correlations as measures of the morphology of structure imply $2-3 \sigma$ deviations of SDSS DR7 luminous red galaxy data from $\Lambda$CDM predictions \cite{Wiegand}. 
Information theory inspired measures of structure \cite{Hosoya,Pandey} have been employed in analysis of large scale catalogues showing convergence of the Shannon entropy to a plateau of \say{almost-homogeneity} at a scale of $\sim 150$Mpc/h for the SDSS DR7 luminous red galaxy sample \cite{SarkarPandeygalaxies} and $\sim 250$Mpc/h for the SDSS DR12 quasar catalogue \cite{SarkarPandeyquasars} respectively. 
Clustering properties can be assessed in combination with knowledge of intrinsic physical properties of the objects examined to form \say{mark correlation functions} \cite{Beisbart}, which might for instance be used to characterise the clustering of galaxies as dependent on their luminosities. 
In this analysis we do not treat intrinsic properties and consider only information on the position of galaxies. We focus on average number counts in spheres as an estimator for the transition to homogeneity. 

Large scale structure of sizes that extend hundreds of Megaparsecs in effective {radii} have been detected and their expectation in the $\Lambda$CDM paradigm debated \cite{ShethDiaferio,ParkSGW,Keenan,Clowes,Park,Marinello,BolejkoOstrowski}. 
Peculiar velocity analysis within the $\Lambda$CDM framework reveal significant coherent motion on scales of several hundreds of Megaparsecs as well \cite{Feindt,Magoulas}, suggesting the need for additional structure in the direction of the Shapley Cluster \cite{Feindt,Colin}. 
While the existence of individual structures of large size do not \emph{a priori} contradict the results of transition to below \say{1\% inhomogeneity} in average number count statistics at scales $\sim 100$Mpc found in \cite{Hogg,Scrimgeour,Laurent}, they do suggest that care must be taken in imposing assumptions about convergence to homogeneity when analysing large scale catalogues. 

The topic of convergence to homogeneity is tightly linked to the topic of self-averaging, in the context of which the issues involving selection effects have been discussed \cite{Labini,LabiniVasilyevBaryshev}. 
Effects of imposing different smoothing scales in the redshift distribution for estimating the survey selection function have been investigated in \cite{CabreGaztanaga}, and the impact of the selection function has also been discussed in relation to Shannon entropy studies \cite{SarkarPandeygalaxies}. 
The systematic effects from modelling the radial selection function directly from the survey has been analysed in the context of $\Lambda$CDM mock catalogues by splitting the mock catalogues in subsamples and using these to estimate the underlying selection function from which the mock catalogues are generated \cite{Mattia}. 

In the present paper we aim at understanding how the imposing of integral constraints on the model survey selection function are expected to bias results for generic matter distributions, focusing on the average number count in spheres (2-point correlation function \cite{PeeblesTheory}) as a homogeneity scale estimator. 
We formalise how typical choices of selection functions can be viewed as equivalent to assuming convergence to homogeneity at the largest scales of the survey. 
To obtain insight through analytic expressions, we consider a simple toy model of an oscillating and asymptotically homogeneous galaxy density distribution. 

The incompleteness of spheres and the use of (homogeneous) artificial catalogues in some analyses to fill survey gaps \cite{Scrimgeour,Laurent} might introduce additional bias towards homogeneity in the estimates, but will be ignored in the present analysis where only spheres fully contained in the survey are considered in the toy model setup. 
Other sources of errors involved in number count statistics are tracers as biased probes of the underlying matter distribution \cite{Desjacques,White} and galaxy evolution \cite{Peng}. 
The effect of finite sampling and finite resolution in number count analysis on the inferred fractal properties of idealised distributions has been analysed in \cite{Poissonfractal,Boxcounting}. 
In the present analysis we consider corrections for selection effects separately from other potential biases. 
When discussing number count statistics we shall refer to galaxies as the counted objects, but our analysis carries over to other tracers.

In section \ref{sec:nfs} we introduce number count based descriptive measures of inhomogeneity and their correction of survey selection effects. 
In section \ref{sec:biasmodel} we consider integral constraints and requirements that are typically imposed for modelling survey selection functions of galaxy surveys. 
Finally, in section \ref{sec:example}, we consider a toy model example where a notion of asymptotic homogeneity is present. We analyse how different estimation procedures of the selection function introduce bias with respect to the \say{bare} number counts. 
We conclude in section \ref{sec:conclusion}. 

\section{Number counts, fractal dimensions, and scales of statistical homogeneity}
\label{sec:nfs}
Here we review measures of inhomogeneity of the matter distribution. 
The number count statistics formulated are analogous to the 2-point correlation function as described in e.g. \cite{PeeblesTheory}. However, here we consider the possibility of a curved space-time and do not restrict the underlying point process from which the galaxies are drawn. The average number count definitions described here can be viewed as purely descriptive measures of the galaxy density distribution within a generic space-time.  
In section \ref{sec:numbercount} we consider \say{bare} number counts as measures of inhomogeneity where selection effects are absent, and in section \ref{sec:biasnumbercounts} we consider the corresponding measures as corrected for selection effects. 

\subsection{Motivation and unnormalised number counts}
\label{sec:numbercount}
We are interested in the distribution of galaxies over a spatial domain equipped with a metric, and to quantify the degree of homogeneity of the distribution at various scales. 
Let us first consider a perfectly homogeneous distribution of galaxies within an appropriate continuity approximation, such that the number count within a given volume $V$ is given by 
\begin{align} \label{eq:homN}
N(V)&= \eta \, V \,  \qquad \textrm{(homogeneous case)} \, ,
\end{align}
where $\eta$ is the constant density of galaxies. 
For a domain equipped with an Euclidean metric, the volume of a sub-domain enclosed by a constant radius $r$ from a center is given by $V(r) = 4 \pi r^3/3$. 
In this case 
\begin{align} \label{eq:homNr}
N(<r)&\equiv N(V(r)) = \frac{4}{3} \eta  \pi r^3 \qquad \textrm{(homogeneous and Euclidean case)} \, .
\end{align}
We might consider a generalisation of the number count (\ref{eq:homNr}) at a point\footnote{In the following the central point $\bm x$ of number counts is taken to be an arbitrary point on the spatial domain. However, when considering point particles in a discretised setting, number counts are most often performed in domains with center coinciding with the position of a galaxy.} on the spatial domain represented by coordinates $\bm x = (x^1, x^2,x^3)$ for domains with generic distributions of matter and general geometry
\begin{align} \label{eq:Nr}
N(\left. <r \right| \bm x)&\equiv \int \dd V_{\bm y}  \, \rho(\bm y) \, \Theta(r - d(\bm x, \bm y)) \, , 
\end{align}
where $d(\bm x,\bm y)$ is the shortest distance from the central point $\bm x$ to the point $\bm y$ defined by the metric tensor $g_{ij}$ (in conventional analysis the distance $d(\bm x,\bm y)$ is computed by employing a hypothesised large scale metric valid for assigning appropriate distances on average for many galaxy pairs), $\rho(\bm y)$ is the galaxy density at the point $\bm y$, $\Theta$ is the unit step function, and $dV_{\bm y} = d^3 y \sqrt{g}$ is the volume 3-form as measured by the metric. 
The function (\ref{eq:Nr}) has dependence on the choice of center $\bm x$ in general. 
{We say that the density distribution exhibits asymptotic homogeneity if there exist a finite constant $\eta \neq 0$ such that we can define the \emph{normalised number count} 
\begin{align} \label{eq:normalisedn}
\mathcal{N}(\left. <r \right| \bm x) \equiv \frac{N(\left. <r \right| \bm x) }{N^{\text{hom}}(\left. <r \right| \bm x)}   \, ,
\end{align}
where the \say{background} homogeneous number count $N^{\text{hom}}(\left. <r \right| \bm x)$ is defined as 
\begin{align} \label{eq:volume}
N^{\text{hom}}(\left. <r \right| \bm x) \equiv \eta \, V( r , \bm x)  \, , \qquad V( r , \bm x) \equiv \int \dd V_{\bm y}  \, \Theta(r - d(\bm x, \bm y))   \, , 
\end{align}
and where $\mathcal{N}(\left. <r \right| \bm x)$ satisfies \cite{Labini} 
\begin{align} \label{eq:assympthom}
\lim_{r \to \infty} \mathcal{N}(\left. <r \right| \bm x) = 1 \quad \forall \, \bm x \, .
\end{align}
We denote $\eta$ the mean galaxy density of the distribution. }
We can define the galaxy-weighted\footnote{{This galaxy-weighted averaging operation definition differs from the corresponding volume-weighted average $\frac{1}{V} \int \dd V_{\bm x} \, N(\left. <r \right| \bm x)$, where $V\equiv  \int \dd V_{\bm x}$, by a weighting factor equal to the galaxy density $\rho(\bm x)$. When $\rho(\bm x)$ is a realisation of a point process, the galaxy-weighted integral becomes the sum of number counts as centered on each galaxy.}} average number count 
\begin{align} \label{eq:Nrint}
N(<r)&\equiv \frac{1}{N} \int \dd V_{\bm x} \, \rho(\bm x) N(\left. <r \right| \bm x)= \frac{1}{N}  \int \dd V_{\bm x}  \dd V_{\bm y}  \, \rho(\bm x) \rho(\bm y) \, \Theta(r - d(\bm x,\bm y)) \, ,
\end{align}
where $N \equiv  \int \dd V_{\bm x} \, \rho(\bm x)$. 
One might define higher order moments to further characterise $N(\left. <r \right| \bm x)$ and corresponding Minkowski-Bouligand dimensions \cite{Yadav} to characterise the scaling behaviour of the moments with $r$. In this analysis we shall analyse the average number count  (\ref{eq:Nrint}) only. 
If (\ref{eq:assympthom}) holds then, assuming commutation of the averaging operation (\ref{eq:Nrint}) and the limit (\ref{eq:assympthom}), {we have that the average normalised number count    
\begin{align} \label{eq:Ncurlydef}
\mathcal{N}( <r ) &\equiv \frac{N( <r ) }{N^{\text{hom}}( <r )}   \, , \qquad  N^{\text{hom}}( <r ) \equiv \eta V( r )    \,  , \qquad   V( r ) \equiv  \frac{1}{N} \int   \dd V_{\bm x}  \, \rho( \bm x) V( r , \bm x)  \,
\end{align}
satisfy 
\begin{align} \label{eq:assympthomav}
\lim_{r \to \infty} \mathcal{N}(<r) = 1   \, .
\end{align} } 
To characterise the scaling behaviour of the number counts we can define the so-called correlation dimension 
\begin{align} \label{eq:FractalDim1}
D_2(\left.r \right| \bm x ) &\equiv \frac{d \ln   N(\left. <r \right| \bm x) }{d  \ln  r } =  \frac{d \ln   \mathcal{N}(\left. <r \right| \bm x) }{d  \ln  r } +  \frac{d \ln   N^{\text{hom}}(\left. <r \right| \bm x) }{d  \ln  r }  \, 
\end{align}
for the number count centered on a single point or 
\begin{align} \label{eq:FractalDim}
D_2(r ) &\equiv \frac{d \ln   N( <r ) }{d  \ln  r }  =  \frac{d \ln   \mathcal{N}(<r ) }{d  \ln  r } +  \frac{d \ln   N^{\text{hom}}(<r) }{d  \ln  r }  \, ,
\end{align}
for the average number count (\ref{eq:Nrint}). 
Deviance of (\ref{eq:FractalDim1}) and (\ref{eq:FractalDim}) from the value $3$ might be due to either non-zero curvature, a flawed radial measure, observational biases, deviance from homogeneity of the galaxy distribution at the given scale, or a combination of these\footnote{These effects might in principle annihilate, such that $D_2(r) \approx 3$ for some range of radii $r$ without the distribution being (close to) homogeneous on these scales. However, this would require chance cancelation of the mentioned effects.}. 

\begin{figure}[!htb]
\centering
\includegraphics[width=0.75 \textwidth]{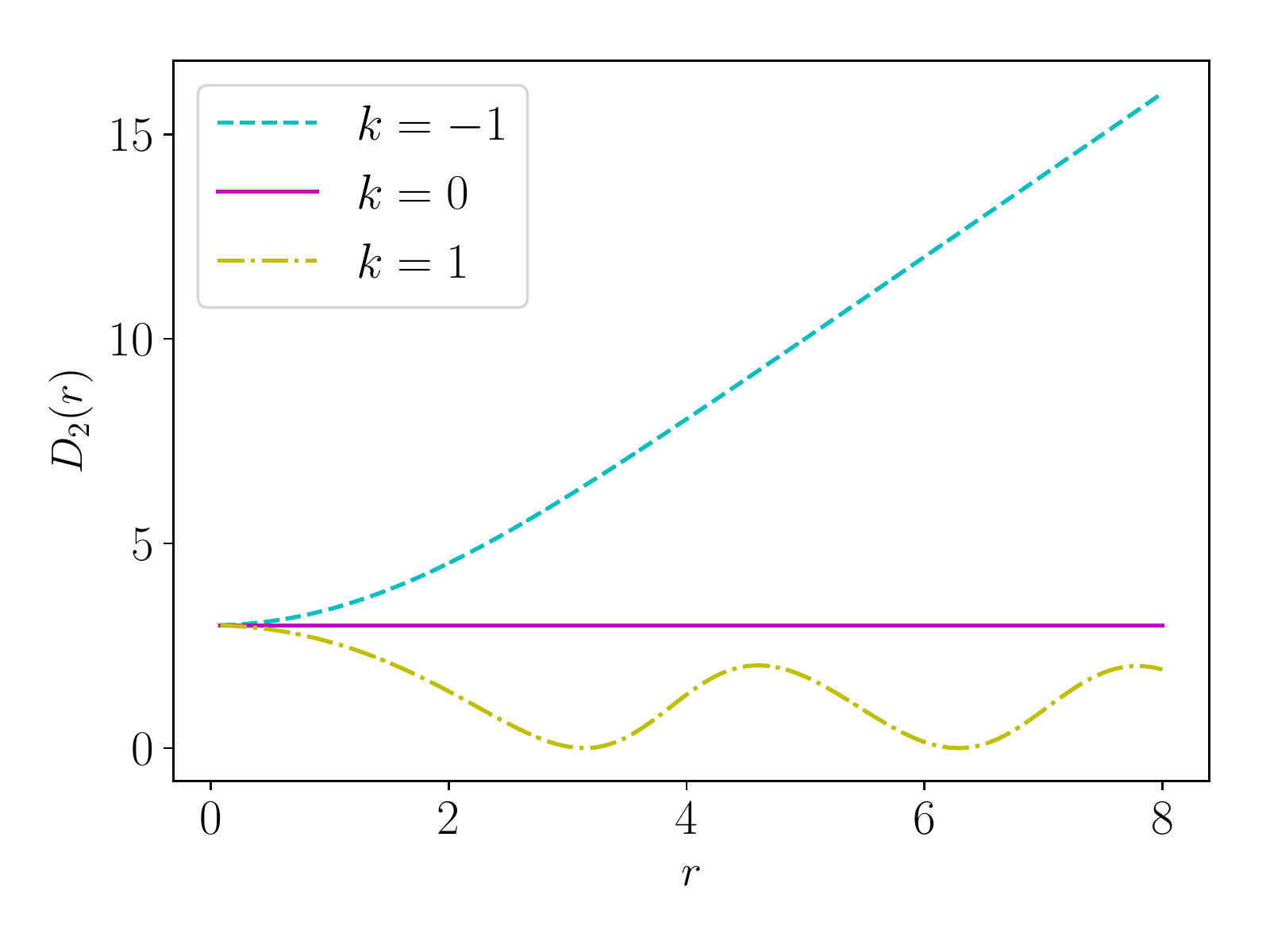}
\caption{The correlation dimension $D_2(r)$ for the open ($k=-1$), flat ($k=0$), and closed ($k=1$) FLRW solutions. 
}
\label{fig:FLRWD2}
\end{figure}

When the Euclidean condition is satisfied such that $V \propto r^3$ it follows that the condition for asymptotic homogeneity (\ref{eq:assympthom}) is equivalent to $D_2(\left.r \right| \bm x ) = 3$ for $r \rightarrow \infty$. 
However for non-Euclidean spatial sections $D_2(\left.r \right| \bm x )$ need not converge to $3$ or even to a constant. 
For an FLRW spacetime with curvature length scale $\sqrt{\abs{k}}^{-1}$ and homogeneous galaxy density over the canonical spatial sections\footnote{The volume of a sphere embedded in an FLRW spatial section can be computed from the line element $d\Sigma^2 \equiv dr^2 + S^2_{k}(r) d\Omega^2$, where $d\Omega^2 = d\theta^2 + \sin^2(\theta) d\phi^2$ is the angular element on the unit sphere, $S_{k}(r) \equiv \sqrt{-k}^{-1} \sinh(\sqrt{-k} r)$ gives the adapted area measure on a sphere of radius $r$. The volume within a sphere of radius $r$ reads $V( r , \bm x) = V(r) = \int_{0}^{r} \dd r' \int \dd \Omega S^2_{k}(r')  = 4\pi  \sqrt{-k}^{-3} ( \frac{\sinh\left(2 \sqrt{-k} r\right)}{4} -\frac{ \sqrt{-k} r}{2} )$, which reduces to $4\pi r^3/3$ for the spatially flat case $k=0$.} we have in the limit of small radii as compared to the curvature scale
\begin{align} \label{eq:FractalDimFLRWsmallr}
D_2(\left.r \right| \bm x ) &= D_2(r ) = 3\left(1 - \frac{2}{15}  \, kr^2\right) \, , \qquad  \sqrt{\abs{k}}r \ll 1  \qquad   \text{(FLRW)} \, , 
\end{align}
where $r$ is the proper radius of the sphere. 
In the limit of large radii as compared to the curvature scale we have 
\begin{align} \label{eq:FractalDimFLRWlarger}
D_2(\left.r \right| \bm x ) &= D_2(r ) = 2  \sqrt{-k} r \, , \qquad  \sqrt{-k}r \gg 1  \qquad   (\text{FLRW} , \; k<0) \,   \\
D_2(\left.r \right| \bm x ) &= D_2(r ) = 2 \sin^2\left(  \sqrt{k} r\right)  \, , \qquad  \sqrt{k} r \gg 1  \qquad   (\text{FLRW} , \; k>0) \, .
\end{align} 
{The exact solutions for the correlation dimension $D_2(r)$ for the open, flat, and closed FLRW solutions are shown in figure \ref{fig:FLRWD2} in units of $\abs{k}=1$ for the curved solutions. 
The open FLRW universe model shows growth of $D_2(r)$, first of approximate quadratic form and then linear form in $r$. 
The closed universe of circumference $\sqrt{k} r = 2\pi$ exhibits oscillations of $D_2(r)$ with local minima at radii $\sqrt{k} r = \pi n$ corresponding to going $n/2$ great circles around the model universe.}   

The FLRW case study exemplifies the importance of curvature in unnormalised number count statistics. 
In order to separate curvature degrees of freedom from the considerations of large scale homogeneity of the galaxy density field, we can consider the convergence of the normalised number counts 
$\mathcal{N}(\left. <r \right| \bm x)$ and  $\mathcal{N}( <r)$.  It follows from (\ref{eq:assympthom}) and (\ref{eq:assympthomav})
\begin{align} \label{eq:assympthomdimensions}
\lim_{r \to \infty} \frac{d \ln   \mathcal{N}(\left. <r \right| \bm x) }{d  \ln  r } = 0 \quad \forall \, \bm x \, ,  \qquad  \lim_{r \to \infty} \frac{d \ln   \mathcal{N}(<r) }{d  \ln  r } = 0
\end{align}
for an asymptotically homogeneous distribution irrespective of the geometry of the spatial section. 

Convergence of $\mathcal{N}(\left. <r \right| \bm x)$ and $\mathcal{N}(<r )$ are insensitive to large scale geometry, because of the normalisation with the volume measure{, and their derivatives per construction tend to zero in the limit of infinite radii. Thus, the normalised number counts $\mathcal{N}(\left. <r \right| \bm x)$, $\mathcal{N}(<r )$ and the associated correlation dimensions in some sense provide more robust measures of inhomogeneity than the unnormalised quantities.}  
Of course, consideration of $\mathcal{N}(\left. <r \right| \bm x)$ and $\mathcal{N}(<r )$ instead of the unnormalised number counts merely pushes the problem of the degeneracy between volume measure and density in the number count to the problem of estimating the background homogeneous number counts $N^{\text{hom}}(\left. <r \right| \bm x)$ and $N^{\text{hom}}(<r)$, which requires assumption on the asymptotic properties of the spatial domain together with knowledge of the volume measure. 

\subsection{Selection effects and normalisation of number counts}
\label{sec:biasnumbercounts}
Catalogues of galaxies cover finite regions of space and are never a perfect sample of the true distribution of galaxies, but subject to selection effects due to, e.g., detector limitations and foreground noise. 
We can incorporate the information about selection effects of a given survey in a selection function $W(\bm x)$ weighting the true underlying density $\rho(\bm x)$ of galaxies at each point, such that the observed number counts within a geodesic radius $r$ of a point $\bm x$ are given by
\begin{align} \label{eq:Nrselection}
\hat{N}(\left. <r \right| \bm x)&\equiv \int \dd V_{\bm y}  \, \hat{\rho}(\bm y) \, \Theta(r - d(\bm x, \bm y)) \, ,  \qquad  \hat{\rho}(\bm y) \equiv  W(\bm y) \, \rho(\bm y)
\end{align}
where $\hat{\rho}(\bm x)$ is the observed galaxy density. 
The number count (\ref{eq:Nrselection}) is in practice defined for sphere centers $\bm x$ and for radii $r$ such that the domain of integration is within the survey volume where $W(\bm x)>0$ almost everywhere. 
When the underlying distribution obeys convergence towards homogeneity (\ref{eq:assympthom}), we define the observed analogue of the homogeneous number count (\ref{eq:volume})---assuming knowledge of the selection function $W(\bm x)$ and the mean density of the distribution $\eta$---as follows 
\begin{align} \label{eq:volumeselection}
\hat{N}^{\text{hom}}(\left. <r \right| \bm x) \equiv \int \dd V_{\bm y}  \, \eta \, W(\bm y)  \, \Theta(r - d(\bm x, \bm y)) = \eta \, \braket{W}_{r, \bm x} V( r , \bm x)  \, , 
\end{align}
where 
\begin{align} \label{eq:Wvolumeav}
 \braket{W}_{r, \bm x}   \equiv \frac{\int \dd V_{\bm y}  \, W(\bm y)  \, \Theta(r - d(\bm x, \bm y)) }{ V( r , \bm x)  }
\end{align}
is the volume average of the selection function over the domain specified by $r$ and $\bm x$. 
We can define an observed normalised number count from (\ref{eq:Nrselection}) and (\ref{eq:volumeselection}) as 
\begin{align} \label{eq:Nnormselection}
\hat{\mathcal{N}}(\left. <r \right| \bm x)  \equiv \frac{\hat{N}(\left. <r \right| \bm x) }{\hat{N}^{\text{hom}}(\left. <r \right| \bm x) } \, . 
\end{align}
Note that (\ref{eq:Nnormselection}) is not guaranteed to converge to unity at the scale of the survey, even though (\ref{eq:assympthom}) holds true asymptotically for the true distribution. 
$\hat{\mathcal{N}}(\left. <r \right| \bm x)$ need also not in general be a good estimate of $\mathcal{N}(\left. <r \right| \bm x)$ at any given scale $r$. 
When $W(\bm y)$ and $\rho(\bm y)$ are uncorrelated in volume such that (\ref{eq:Nrselection}) reduces to 
\begin{align} \label{eq:Nrselectionuncorr}
\hat{N}(\left. <r \right| \bm x)&=   \braket{W}_{r, \bm x}   \int \dd V_{\bm y}   \, \rho(\bm y) \, \Theta(r - d(\bm x, \bm y))  \qquad \quad \text{(uncorrelated case)}  \, , 
\end{align}
we have that $\hat{\mathcal{N}}(\left. <r \right| \bm x) =\mathcal{N}(\left. <r \right| \bm x)$. The uncorrelated approximation (\ref{eq:Nrselectionuncorr}) is expected to be good when the selection function of galaxies is not determined by the underlying galaxy density distribution itself, i.e., that the probability of observing a given galaxy is independent of the number density of galaxies in its vicinity. 

We define the galaxy-weighted empirical number count as follows  
\begin{align} \label{eq:Nrselectionav}
\hat{N}(<r)&\equiv \frac{1}{\hat{N}_{\mathcal{D}}} \int_{\mathcal{D}} \dd V_{\bm x}  \, \hat{\rho}(\bm x) \, \hat{N}(\left. <r \right| \bm x)
\end{align}
where $\hat{N}_{\mathcal{D}} \equiv  \int_{\mathcal{D}} \dd V_{\bm x}  \, \hat{\rho}(\bm x)$, and where the domain of integration $\mathcal{D}$ might be a subset of the volume covered by the survey in order to exclude spheres not fully contained in the survey. 
Similarly we define the galaxy-weighted empirical homogeneous number count as 
\begin{align} \label{eq:Nrselectionavhom}
N_W^{\text{hom}}(<r) &\equiv \frac{1}{\hat{N}_{\mathcal{D}}} \int_{\mathcal{D}} \dd V_{\bm x}  \, \hat{\rho}(\bm x) \, \hat{N}^{\text{hom}}(\left. <r \right| \bm x) \, , 
\end{align} 
from which we can define the empirical normalised average number count\footnote{This definition is analogous to the estimator of the 2-point correlation function as suggested by Davis and Peebles \cite{DPestimator}. For a review of alternative estimators and their efficiency within $\Lambda$CDM, see \cite{Kerscher}. } 
\begin{align} \label{eq:NnormavselectionW}
\mathcal{N}_W( <r)  \equiv \frac{\hat{N}(<r ) }{N_W^{\text{hom}}(<r) } \, . 
\end{align}
While (\ref{eq:Nrselection}) and (\ref{eq:Nrselectionav}) can be determined directly from the catalogue through the observed coordinate adapted galaxy density $\sqrt{g}(\bm x) \hat{\rho}(\bm x)$---along with assumption about the metric tensor for computing distances between the galaxies---the empirical homogeneous counts (\ref{eq:volumeselection}) and (\ref{eq:Nrselectionavhom}) require assumptions about the selection function $W(\bm x)$ and asymptotic mean galaxy density $\eta$, which are not directly observed. 
$\eta$ is degenerate with the overall normalisation of $W(\bm x)$. 
We denote estimates of the selection function $\hat{W}(\bm x)$ and define the corresponding estimated number count
\begin{align} \label{eq:Nnormavselection}
\hat{\mathcal{N}}( <r)  \equiv \frac{\hat{N}(<r ) }{N_{\hat{W}}^{\text{hom}}(<r) } \, 
\end{align}
relevant for observation.

While the overall normalisation of $\hat{W}(\bm x)$ amounts to a scaling of normalised number counts (\ref{eq:Nnormselection}) and (\ref{eq:Nnormavselection}), different assumptions about the shape of $\hat{W}(\bm x)$ can result in entirely different estimates of scaled number counts.  
As an example we might assume $\hat{W}(\bm x) = const.$, which represents the assumption of a perfectly sampled galaxy density (no selection effects). As another limiting example we can consider $\hat{W}(\bm x) \propto \hat{\rho}(\bm x)$, i.e., assuming that all structure in the observed catalogue is due to selection effects, in which case the scaled number counts (\ref{eq:Nnormselection}) and (\ref{eq:Nnormavselection}) are constant in $r$ by construction\footnote{It is likely not necessary to choose $\eta \hat{W}(\bm x) = \hat{\rho}(\bm x)$ to obtain $\hat{\mathcal{N}}( <r) =1$. Since $\eta \hat{W}(\bm x)$ is a function of three variables with no a priori constraints (other than the requirement of being a positive function) we expect there to be an infinity of functions $\hat{W}(\bm x)$ satisfying the one parameter family of constraints $\hat{\mathcal{N}}( <r) =1$.}.

\section{Modelling of the survey selection function}
\label{sec:biasmodel}
Corrections for survey selection effects through an assumed selection function $\hat{W}(\bm x)$ as described in section \ref{sec:biasnumbercounts} might be necessary when the volume covered by the survey is not fairly sampled. 
The reliability of the results in this case rely on the ability to properly estimate the selection effects. 
The selection effects due to luminosity selection cuts, detector limitations, and foreground noise are hard to model independently, and crude assumptions, which are implicitly assuming large scale convergence to homogeneity, are often used in concrete analysis. 
Here we discuss different modelling assumptions for selection functions relevant for modern large scale surveys \cite{Reid,Myers:2015hpw,DESI}. 
{In section \ref{sec:largescalehom} we discuss the assumption of convergence of the mean density within the survey volume to the mean density of a full spatial section of the Universe. 
In section \ref{sec:decoupling} we discuss the assumption about decoupling of angular and radial selection effects in galaxy surveys. Finally in section \ref{sec:noinhomrad} and section \ref{sec:noinhom} we consider the assumption of cross sectional homogeneity over angular and radial scales of the survey domain. 
}

\subsection{Large scale homogeneity}
\label{sec:largescalehom}
Let us consider a distribution of objects that obeys spatial homogeneity in the form of (\ref{eq:assympthom}). 
Let us further assume that the survey spans a volume that is large enough to accurately sample the mean density of the survey, such that the limit (\ref{eq:assympthom}) holds approximately within volumes comparable to the survey such that 
\begin{align} \label{eq:assympthomD}
N & \equiv \int_{\text{survey}} \dd V_{\bm y}  \, \rho(\bm y) \approx \eta \int_{\text{survey}} \dd V_{\bm y}  \, 
\end{align}
where integration is over the volume covered by the survey. 
Requiring the same limiting behaviour for the estimated number count implies 
\begin{align}\label{eq:intconstraint} 
\hat{N} & \equiv \int_{\text{survey}} \dd V_{\bm y}  \, \hat{\rho}(\bm y) \approx   \eta \int_{\text{survey}} \dd V_{\bm y}  \, \hat{W}(\bm y) \, .
\end{align}
This integral constraint fixes the normalisation of the selection function in terms of the unknown mean galaxy density $\eta$ of the distribution. 

\subsection{Decoupling of radial and angular dependence}
\label{sec:decoupling}
Let us consider observer adapted spherical coordinates $\bm x = (\rad,\theta,\phi)$. Typically $\theta,\phi$ will correspond to the angular position on the sky of an observer, and $\rad$ will coincide with the redshift function as measured by the observer, assuming that redshift is indeed a valid radial coordinate\footnote{This is the case for e.g. expanding FLRW models or LTB models. In general we expect the redshift function defined on a null bundle with respect to a class of observers to be multivalued. This might physically happen when e.g. the null rays pass collapsing regions of space. One might circumvent such issues by employing a background or template metric valid on large scales where collapse of structure might be ignored. In practice, large scale galaxy catalogues are typically analysed employing large scale geometric measures.}. 
We might make the assumption that the selection function is multiplicative separable in the coordinate basis $\bm x = (\rad,\theta,\phi)$ such that\footnote{The number count in a small observer adapted coordinate box $\Delta \rad \Delta \theta \Delta \phi$ is given by $\Delta \hat{N}(\rad,\theta,\phi) =  \hat{\rho} \sqrt{g} \, \Delta \rad \Delta \theta \Delta \phi =  \rho \, \hat{W} \sqrt{g} \, \Delta \rad \Delta \theta \Delta \phi$. It is therefore natural to impose constraints on $\eta \hat{W} \sqrt{g}$ in the coordinate basis $\bm x = (\rad,\theta,\phi)$ as it defines the selection function within a coordinate volume that is naturally and model independently defined by the observer.} 
\begin{align} \label{eq:rhoRseparable}
\sqrt{g}\, \hat{W}(\rad,\theta,\phi)\, = A(\rad)\, B(\theta,\phi) \, .
\end{align}
Such a model assumption might be justified in cases where the individual selection biases affecting the survey are dependent on either redshift or angles. As an example, angular limitation of coverage of telescopes affect the galaxy density distribution solely in the angular direction. 
Malmquist type biases are a function of the cross section of the null bundle at emission and observation respectively, and might be thought of as a function of radial distance for models with a notion of statistical spherical symmetry. 
In \cite{Ross} it was argued on empirical grounds for a multiplicative separable survey selection function, and multiplicative separability is a typical invoked assumption for galaxy surveys. 

\subsection{Cross sectional angular homogeneity}
\label{sec:noinhomrad}
As in section \ref{sec:largescalehom} we consider an underlying distribution that obeys the spatial homogeneity condition (\ref{eq:assympthom}). 
We now consider situations where the mean galaxy density of the underlying distribution is well sampled over angular cross sections of the survey, such that 
\begin{align}  \label{eq:integralRho2theory1} 
\bar{\rho}(\rad)  \equiv \int_{\text{survey}} \dd \theta \dd \phi \, \sqrt{g} \rho(\rad,\theta,\phi) \approx \eta \int_{\text{survey}} \dd \theta \dd \phi \, \sqrt{g}(\rad,\theta,\phi)   \, .
\end{align}
Requiring the estimated normalised number counts to obey the same convergence as the underlying number counts results in the following cross sectional relation: 
\begin{align}  \label{eq:integralRho21} 
\hat{\bar{\rho}}(\rad)  & \equiv \int_{\text{survey}} \dd \theta \dd \phi \, \sqrt{g} \hat{\rho}(\rad,\theta,\phi) \approx \eta \int_{\text{survey}} \dd \theta \dd \phi \, \sqrt{g} \hat{W}(\rad,\theta,\phi)  \, .
\end{align}
The overall integral constraint (\ref{eq:intconstraint}) is automatically satisfied when (\ref{eq:integralRho21}) is satisfied. 
The selection function ansatz (\ref{eq:integralRho21}) is in practice applied to construct random galaxy catalogues for correcting for selection effects for the SDSS-III Baryon Oscillation Spectroscopic Survey \cite{Reid,Ata}.\footnote{There is in practice a small level of stochasticity related to the random assignment of redshifts to the particles of the random catalogue implying a slight difference between the galaxy catalogue and random catalogue radial distribution.} 
Additional assumptions must be invoked to fully determine the selection function $\hat{W}(\rad,\theta,\phi)$. This is typically done by assuming decoupling of angular and radial dependence \ref{eq:rhoRseparable}, such that $A(\rad) \propto \hat{\bar{\rho}}(\rad)$, and determining $B(\theta,\phi)$ through assessing the level of completeness in detection in angular patches of the sky \cite{Reid}. Alternatively $B(\theta,\phi)$ might be determined from assuming cross sectional radial homogeneity in the same spirit as the cross sectional angular homogeneity imposed in the present section (see the subsection below).

\subsection{Cross sectional radial and angular homogeneity}
\label{sec:noinhom}
We now consider situations where the mean density of the underlying distribution of galaxies is well sampled in thin lines and sheets of scales corresponding to the radial and angular cross sections of the survey, such that 
\begin{align} \label{eq:assympthomDranglestheory}
\bar{\rho}(\theta,\phi) & \equiv \int_{\text{survey}} \dd \rad  \, \sqrt{g} \rho(\rad,\theta,\phi) \approx \eta \int_{\text{survey}} \dd \rad   \, \sqrt{g}(\rad,\theta,\phi)    \, , \\
\bar{\rho}(\rad)  & \equiv \int_{\text{survey}} \dd \theta \dd \phi \, \sqrt{g} \rho(\rad,\theta,\phi) \approx \eta \int_{\text{survey}} \dd \theta \dd \phi \, \sqrt{g}(\rad,\theta,\phi)   \,  \label{eq:integralRho2theory} 
\end{align}
in adapted spherical coordinates\footnote{The formulated integral constraints are independent on the choice of radial coordinate and angular coordinate system respectively. Thus, we might choose any bijective mapping $\rad \mapsto \rad '(\rad )$ and $(\theta,\phi)\mapsto (\theta'(\theta,\phi),\phi'(\theta,\phi))$ for formulating the integral constraints.} $(\rad,\theta,\phi)$ to the observer holds. 
Requiring the estimated normalised number counts to obey the same convergence as the underlying number counts results in the following cross sectional relations:  
\begin{align} \label{eq:integralRho1}
\hat{\bar{\rho}}(\theta,\phi) & \equiv \int_{\text{survey}} \dd \rad  \, \sqrt{g} \hat{\rho}(\rad,\theta,\phi) \approx \eta \int_{\text{survey}} \dd \rad   \, \sqrt{g} \hat{W}(\rad,\theta,\phi)   \, , \\
\hat{\bar{\rho}}(\rad)  & \equiv \int_{\text{survey}} \dd \theta \dd \phi \, \sqrt{g} \hat{\rho}(\rad,\theta,\phi) \approx \eta \int_{\text{survey}} \dd \theta \dd \phi \, \sqrt{g} \hat{W}(\rad,\theta,\phi)  \, . \label{eq:integralRho2} 
\end{align}
The overall integral constraint (\ref{eq:intconstraint}) is automatically satisfied when either (\ref{eq:integralRho1}) or (\ref{eq:integralRho2}) is satisfied. 
The assumptions about separate angular and radial convergence are in general not trivial to satisfy even though the overall integral constraint (\ref{eq:intconstraint}) might be satisfied to a good approximation for the given survey volume. 
The details about the survey geometry are important for the accuracy of the approximations (\ref{eq:assympthomDranglestheory}) and (\ref{eq:integralRho2theory}).

Combining the constraints (\ref{eq:integralRho1}) and (\ref{eq:integralRho2})---replacing the approximation with strict equality---with (\ref{eq:rhoRseparable}) fully specifies $\sqrt{g}\hat{W}(\rad,\theta,\phi)$ 
\begin{align} \label{eq:rhoRseparableConstrained}
\sqrt{g}\, \eta  \hat{W}(\rad,\theta,\phi) &= \frac{1}{\hat{N}} \hat{\bar{\rho}}(\rad)\, \hat{\bar{\rho}}(\theta,\phi) \,  ,
\end{align}
{which can be derived by plugging the ansatz (\ref{eq:rhoRseparable}) into the right hand sides of (\ref{eq:integralRho1}) and (\ref{eq:integralRho2}) to determine the functions $A(\rad)$ and $B(\theta,\phi)$.}
This type of selection function was proposed in \cite{Burden} {(see their section 3.1)} for the DESI experiment \cite{DESI} to correct for fiber collision\footnote{The finite distance between fibers in fiber-fed spectroscopic surveys, causes failure in assigning a fiber to each galaxy for measuring its redshift and hence limits the angular resolution of the survey. See, e.g, \cite{Guo} for a review on fiber collisons.}, and investigated in \cite{Burden,Mattia} using $\Lambda$CDM mock catalogues. 

\section{Example: A statistically homogeneous galaxy distribution and estimates of the survey selection function}\label{sec:example}
We now consider an idealised example of a galaxy density distribution over a flat spatial section, where there is a notion of convergence towards homogeneity on the largest scales, but where smaller scales exhibit non-linear fluctuations around the mean density field. We consider a situation where the galaxy density is perfectly sampled within the survey, such that the true selection function is equal to unity within the survey domain. 
In section \ref{sec:exampleproperties} we analyse the properties of the galaxy density distribution and associated number counts and correlation dimensions. 
In section \ref{sec:spherical} we consider number counts as formulated within an spherically shaped survey and discuss the significance of different estimates of the selection function. 
{In section \ref{sec:discussion} we discuss the implications of the results in terms of volume coverage of typical galaxy catalogues and of the sizes of the largest structures in the matter distribution known to date.}

\subsection{The model density distribution}
\label{sec:exampleproperties}
We consider a three dimensional spatial section with adapted Euclidean metric 
\begin{align} \label{eq:metricexampleadapted3}
\dd s^2 = \dd x^2 + \dd y^2 + \dd z^2   \, ,
\end{align}
and take the galaxy density field over the surface to be given as\footnote{The model galaxy density distribution is similar to that considered as a source to Einstein's field equations in \cite{SikoraGlod} for studying backreaction effects in a statistically homogeneous and isotropic spacetime. Since the main objective of this paper is to study selection effects, we shall maintain the Euclidean metric approximation in (\ref{eq:metricexampleadapted3}) for convenience of computation, but note that density fluctuations will in general be associated with corresponding corrections to the Euclidean volume element and distance measures in a realistic universe model with structure.}
\begin{align} \label{eq:densityexample3}
\rho(x,y,z) = \mathcal{K} \left( 1 + \frac{1}{3}\sin\left(Bx + B\overset{o}{x}\right) + \frac{1}{3}\sin\left(By + B\overset{o}{y}\right)  + \frac{1}{3}\sin\left(Bz + B\overset{o}{z}\right) \right) \, , 
\end{align}
{where $B$ is the frequency of the oscillation of the density field, and $\{\overset{o}{x},\overset{o}{y},\overset{o}{z}\}$ represents a translation of the center of the spatial coordinate system $\{x,y,z\}$.}   
With this definition of galaxy density, there is clearly a notion of statistical homogeneity and isotropy, while the density field is locally non-linear and oscillates with values between $0$ and $2 \mathcal{K}$ {(see figure \ref{subfig:density_3D_rhocenter1} and figure \ref{subfig:density_3D})}.
We can rewrite (\ref{eq:metricexampleadapted3}) and (\ref{eq:densityexample3}) in spherical coordinates $\{x,y,z\} \rightarrow \{\rad , \theta, \phi \}$, given by $\{x,y,z\}  = \{\rad \sin(\theta) \cos(\phi), \,  \rad \sin(\theta) \sin(\phi) , \,  \rad \cos(\theta) \}$ with $r\geq 0$, $\theta \in [0, \pi]$, $\phi \in [-\pi, \pi]$. 
In these coordinates the metric and its determinant reads: 
\begin{align} \label{eq:metricexamplepolar3}
\dd s^2 = \dd \rad^2 + \rad^2 \left( \dd \theta^2    +   \sin^2(\theta)  \dd \phi^2 \right) \, , \qquad g(\rad,\theta,\phi) = \rad^4 \sin^2(\theta)
\end{align}
respectively, and the density takes the form 
\begin{align} \label{eq:densityexamplepolar3}
\rho(\rad,\theta,\phi) = \mathcal{K} \left( 1 + \frac{1}{3}\sin\left(B \rad \sin(\theta)  \cos(\phi) + B \overset{o}{x}   \right) + \frac{1}{3}\sin\left(B \rad \sin(\theta)  \sin(\phi) + B \overset{o}{y}    \right) \right. \nonumber \\
\left. + \frac{1}{3}\sin\left(B \rad \cos(\theta)+ B \overset{o}{z}    \right) \right) \, . 
\end{align}
The number count within a sphere of radius $r$ and as centered at a point $\bm x = (\rad,\theta,\phi)$ can be computed from the definition in (\ref{eq:Nr}) and reads:
\begin{align} \label{eq:Numbercountexample3}
& N\left(\left. <r \right| \bm x  \right) =  \frac{4}{3}\pi \mathcal{K} r^3 \left(  1 +  \frac{k(\bm x ) \, \Bigl( -B r \cos(B r) + \sin(B r) \Bigr) }{  B^3 r^3   }  \right) \, , \quad  k(\bm x ) \equiv  3 (\rho(\bm x ) - \mathcal{K})  \, ,
\end{align}
with asymptotic equivalence
\begin{align} \label{eq:Numbercountexample3lim}
& \lim_{r \to \infty} \frac{N\left(\left. <r \right|  \bm x \right) }{  \mathcal{K} V(r)} = 1 \, , \qquad V(r) \equiv \frac{4}{3}\pi r^3. 
\end{align}
We thus identify the mean galaxy density of the distribution as $\mathcal{K} $ and define the homogeneous background number count {according to} (\ref{eq:assympthom}) and (\ref{eq:assympthomav})
\begin{align} \label{eq:examplehom} 
N^{\text{hom}}\left(\left. <r \right| \bm x  \right) = N^{\text{hom}}(<r) =  \mathcal{K} V(r) \, .
\end{align}
We decompose (\ref{eq:Numbercountexample3}) in terms of $N^{\text{hom}}\left(\left. <r \right| \bm x  \right)$ in the following way
\begin{align} \label{eq:Numbercountexample3decomposed}
N\left(\left. <r \right| \bm x \right) &=   \mathcal{N}\left(\left. <r \right| \bm x \right) N^{\text{hom}}(<r)
\end{align}
with 
\begin{align} \label{eq:exampleinhom}
\mathcal{N}\left(\left. <r \right| \bm x \right) \equiv   1 +  \frac{k(\bm x) \, \Bigl( -B r \cos(B r) + \sin(B r) \Bigr) }{  B^3 r^3   } \, . 
\end{align}
In the special case where the center of the sphere is located at a point of average galaxy density $\mathcal{K}$, i.e., $k(\bm x) = 0$, we have that the number count resembles that of a homogeneous distribution with $N\left(\left. <r \right| \bm x \right) = N^{\text{hom}}(<r)$. Conversely, the larger the density contrast of the center of the domain, the slower the convergence towards homogeneous behaviour of the number count. 

For scales much larger than the period of oscillation, the number count experiences oscillations around the asymptotic homogeneous number count $N^{\text{hom}}(<r)$ with an amplitude that is damped at large scales. 
Using the expansion of $\sin(Br)$ and $\cos(Br)$ we have 
\begin{align} \label{eq:Numbercountexamplelimits3}
N\left(\left. <r \right| \bm x \right) \approx   \frac{4}{3}\pi \mathcal{K} r^3 \left(  1 +  \frac{k(\bm x)}{3} -  \frac{k(\bm x)}{30} (Br)^2 \,  \right)     \qquad  & \text{for} \, Br \ll 1 \, .
\end{align}
For scales much smaller than the period of the oscillation in the galaxy density field, the leading order number count $N\left(\left. <r \right| \bm x \right)$ is equal to that of a homogeneous distribution with density equal to that of the center of the sphere. When the galaxy density at the center of the sphere is zero, the leading order number count is given by a powerlaw dependence $\propto r^5$.

The galaxy-weighted average number count (\ref{eq:Nrint})---obtained by averaging (\ref{eq:Numbercountexample3}) over an infinite domain---reads:
\begin{align} \label{eq:Numbercountavexampleinf}
N\left( <r  \right) = \mathcal{N}\left( <r \right) N^{\text{hom}}(<r) 
\end{align}
with 
\begin{align} \label{eq:Numbercountavexampleinfdef}
\mathcal{N}\left( <r \right)  \equiv 1 + \frac{1}{2} \, \frac{ -B r \cos(B r) + \sin(B r)  }{  B^3 r^3   }   \, , 
\end{align}
where a decomposition similar to that in (\ref{eq:Numbercountexample3decomposed}) has been done. 
Taking the galaxy-weighted average of (\ref{eq:Numbercountexample3}) over an infinite volume thus amounts to making the replacement $k(\bm x) \mapsto 1/2$ in (\ref{eq:Numbercountexample3}).\footnote{The {galaxy-weighted} averaging operation is important for this result. Performing, for instance, a volume-weighted integral over $N\left(\left. <r \right| \bm x \right)$ yields zero on the largest scales.}

We might define a radial scale at which fluctuations of (\ref{eq:Numbercountexample3}) around the homogeneous background count (\ref{eq:examplehom}) converge to levels $\leq \alpha$ for all possible choices of center of the sphere  
\begin{align} \label{eq:homscaleeq}
r_{\text{hom}}^{\mathcal{N}\left(\left. <r \right| \bm x \right) } = \text{max}\left( \{ r \in \mathbb R_{\geq 0} \, : \abs*{ \mathcal{N}\left(\left. <r \right| \tilde{\bm x} \right) - 1 } = \alpha \} \right) \, ,  %
\end{align}
where $\tilde{\bm x}$ is a point of maximal density contrast, i.e., $\abs{k(\tilde{\bm x})} = 3$.

We might also consider the conventional definition of a statistical homogeneity scale from level sets of the mass averaged number count 
\begin{align} \label{eq:homscaleeq2}
r_{\text{hom}}^{\mathcal{N}\left( <r \right) } =  \text{max}\left( \{ r \in \mathbb R_{\geq 0} \, : \abs*{ \mathcal{N}\left( <r \right)  - 1 }  = \alpha \} \right) \, .  
\end{align}
From the leading order behaviour at large $Br$ it can be seen that the scale (\ref{eq:homscaleeq2}) will be smaller than (\ref{eq:homscaleeq}) for the same choice of inhomogeneity level $\alpha$ by a factor of $\sim \sqrt{6}${, illustrating that number counts centered on sufficiently overdense or underdense points converge slower to the background homogeneous count than the average number count.} 

\begin{figure}[!htb]
\centering
\includegraphics[width=0.7 \textwidth]{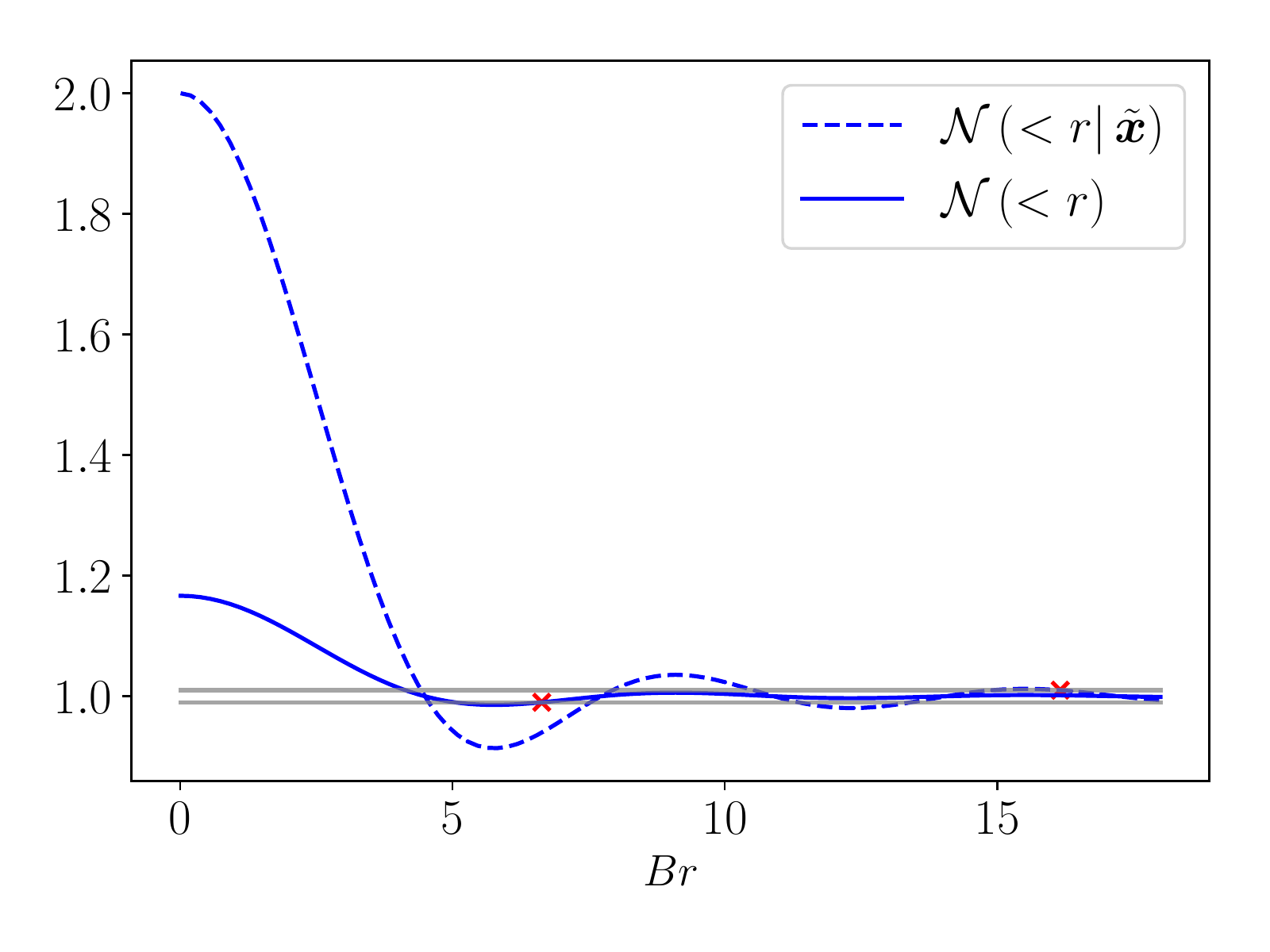}
\caption{{Number counts $\mathcal{N}\left(\left. <r \right| \tilde{\bm x} \right)$ with $k(\tilde{\bm x})=3$ and $\mathcal{N}\left(<r  \right)$ shown as a function of $Br$. The grey horizontal lines illustrate the homogeneity threshold $1 \pm \alpha$ for $\alpha = 0.01$. The red crosses mark the values $B r_{\text{hom}}^{\mathcal{N}\left(\left. <r \right| \bm x \right) }  = 16.16$ and $B r_{\text{hom}}^{\mathcal{N}\left( <r \right) }  = 6.64$ for which the number counts converge below 1\% inhomogeneity. 
}}
\label{fig:NumbercountsTheory}
\end{figure}

{In figure \ref{fig:NumbercountsTheory} the scaled number count $\mathcal{N}\left(\left. <r \right| \tilde{\bm x} \right)$ associated with the toy model density (\ref{eq:densityexample3}) is shown for a central point $\tilde{\bm x}$ of maximal density along with the average scaled number count $\mathcal{N}\left(<r  \right)$. 
The grey horizontal lines depict $1 \pm \alpha$ for the choice of inhomogeneity level $\alpha = 0.01$. The intersection of the $1 \pm \alpha$ lines with the number counts (marked by red crosses) correspond to the solutions to (\ref{eq:homscaleeq}) and (\ref{eq:homscaleeq2}) yielding the homogeneity scales $B r_{\text{hom}}^{\mathcal{N}\left(\left. <r \right| \bm x \right) }  = 16.16$ and $B r_{\text{hom}}^{\mathcal{N}\left( <r \right) }  = 6.64$ respectively. 
It is clear from the solutions to $r_{\text{hom}}^{\mathcal{N}\left(\left. <r \right| \bm x \right) }$ and $r_{\text{hom}}^{\mathcal{N}\left( <r  \right) } $ and from figure \ref{fig:NumbercountsTheory} that there will} be spheres with radius $Br = 6.64$ with number count contrast of well above $\sim 1\%$, while there is a notion of convergence of \say{average} number count contrast below $\sim 1\%$ around such a radial scale. 

\subsection{Survey selection function estimation for a model survey}
\label{sec:spherical}
We now examine the situation where the distribution (\ref{eq:densityexamplepolar3}) is observed over a finite region of the sky.
We consider an idealised situation where the underlying galaxy density distribution is perfectly sampled by the survey catalogue, i.e., $\hat{\rho}(\bm x) = \rho(\bm x)$ within the volume covered by the survey. 
Any correction applied to the galaxy number counts with an estimated selection function $\hat{W}(\bm x) \neq 1$ within the survey will thus represent an effective \emph{introduction} of bias instead of removal of selection effects. 
In the following we employ typical methods for estimating the galaxy selection function presented in section \ref{sec:biasmodel} and assess how these bias homogeneity scale estimates for an spherically shaped survey. 

We define a survey domain covering a region of space by $\rad \in [0, R_{\mathcal{D}}]$, $\theta \in [0, \pi]$, $\phi \in [-\pi, \pi]$.
The coordinates $(\rad,\theta,\phi)$ are taken to be suitable adapted coordinates to an observer located in the center of the domain $\rad=0$. 
Due to the finite size of the domain, the \emph{empirical} galaxy-weighted average number count and corresponding artificial average number counts are constructed such that integration is over spheres that are fully contained in the survey
\begin{align} \label{eq:Numbercountavexample3}
N\left( <r  \right) = \frac{1}{N(R_{\mathcal{D}} - r) } \int_{0}^{R_{\mathcal{D}} - r} \dd \rad \int_{0}^{\pi} \dd \theta \int_{-\pi}^{\pi} \dd \phi \,  \sqrt{g} \rho(\rad,\theta,\phi)  N\left(\left. <r \right| (\rad, \theta, \phi) \right) \, , 
\end{align}
\begin{align} \label{eq:NumbercountavRexample3}
N^{\text{hom}}_{\hat{W}}\left( <r  \right) = \frac{1}{N(R_{\mathcal{D}} - r) } \int_{0}^{R_{\mathcal{D}} - r} \dd \rad \int_{0}^{\pi} \dd \theta \int_{-\pi}^{\pi} \dd \phi \,  \sqrt{g} \rho(\rad,\theta,\phi)  N^{\text{hom}}_{\hat{W}}\left(\left. <r \right| (\rad, \theta, \phi) \right) \, , 
\end{align}
where 
$N(d) \equiv N\left(\left. <d  \right| \bm x =\bm  0 \right)$ 
is the number count in the domain defined by $\rad \in [0, d]$, $\theta \in [0, \pi]$, $\phi \in [-\pi, \pi]$. 
The marginalised radial galaxy density distributions for the survey reads: 
\begin{align} \label{eq:densityexamplepolarr3}
\bar{\rho}(\rad) &\equiv  \int_{0}^{\pi} \dd \theta \int_{-\pi}^{\pi} \dd \phi \, \sqrt{g(\rad,\theta,\phi)}  \rho(\rad,\theta,\phi) = 4 \pi \mathcal{K} \rad^2 \left(  1+ \frac{1}{3} k(\bm 0)  \frac{\sin(B\rad)}{B\rad}  \right) \, .  
\end{align}
For a domain centered on a point of average galaxy density $k(\bm 0)= 0$, the radial galaxy density function (\ref{eq:densityexamplepolarr3}) resemble that of the homogenous density function $\rho(\rad,\theta,\phi) = \mathcal{K}$\footnote{This result is a consequence of the spherical symmetry of the survey imposed in this example. Cutting out a cone like structure, by e.g. restricting the survey volume to lie in the positive halfplane $x\geq 0, y\geq 0, z\geq 0$, would for instance yield correction terms to the homogeneous law.}. 
The marginalised angular galaxy density reads: 
\begin{align} \label{eq:densityexamplepolarphi3}
 \bar{\rho}_{R_{\mathcal{D}}}(\theta, \phi)  &\equiv \int_{0}^{R_{\mathcal{D}}} \dd \rad \, \sqrt{g(\rad,\theta,\phi)}  \rho(\rad,\theta,\phi)  \nonumber   \\
&= \frac{1}{3}  \mathcal{K} R_{\mathcal{D}}^3 \, \sin(\theta) \left(1  + \cos(B \overset{o}{x}) \mathcal{A}(\theta,\phi) + \sin(B \overset{o}{x}) \mathcal{B}(\theta,\phi) + \cos(B \overset{o}{y}) \mathcal{A}(\theta,\frac{\pi}{2} - \phi)   \right. \nonumber  \\
&\left. + \sin(B \overset{o}{y}) \mathcal{B}(\theta,\frac{\pi}{2}  - \phi)   + \cos(B \overset{o}{z}) \mathcal{A}(\frac{\pi}{2}  - \theta,0) + \sin(B \overset{o}{z}) \mathcal{B}(\frac{\pi}{2}  - \theta,0)     \right)  \nonumber \\ 
 \mathcal{A}_{\overset{o}{x}}(\theta,\phi) & \equiv   \frac{-2 + (2 -  B^2 R_{\mathcal{D}}^2 f^2(\theta,\phi) ) \cos(B R_{\mathcal{D}} f(\theta,\phi) ) + 2  B R_{\mathcal{D}} f(\theta,\phi)  \sin(B R_{\mathcal{D}}  f(\theta,\phi)  )}{B^3 R^3_{\mathcal{D}}  f^3(\theta,\phi) }   \nonumber \\
 \mathcal{B}_{\overset{o}{x}}(\theta,\phi) & \equiv \frac{2 B R_{\mathcal{D}} f(\theta,\phi)  \cos(B R_{\mathcal{D}} f(\theta,\phi) ) + (-2 + B^2 R_{\mathcal{D}}^2 f^2(\theta,\phi) ) \sin(B R_{\mathcal{D}} f(\theta,\phi) )}{B^3 R^3_{\mathcal{D}} f^3(\theta,\phi) }   \, ,
\end{align} 
where $f(\theta,\phi) \equiv \sin(\theta) \cos(\phi)$. 

We shall consider different choices of estimated selection functions for normalising the number counts of the survey. 
First we consider the \say{bare} estimated selection function coinciding with the true selection function $\hat{W}^{\text{bare}}(\bm x)=W(\bm x)$,  where $W(\bm x) = 1$ when $\bm x$ is in the survey domain, and  $W(\bm x) = 0$ otherwise. The corresponding homogeneous number count reads: 
\begin{align} \label{eq:NrRdecomposedtrue}
N^{\text{hom} }_{\hat{W}^\text{bare}}(\left. <r \right| \bm x) &= N^{\text{hom}}( <r ) \, ,
\end{align}
where $N^{\text{hom}}(<r)$ is given by (\ref{eq:examplehom}). 
The bare homogeneous number count (\ref{eq:NrRdecomposedtrue}) does in general not satisfy the convergence requirement on the scale of the survey as described in section \ref{sec:largescalehom} because of physical structure in the model galaxy distribution and the finite size of the model survey. 
We might consider the simple constant rescaling of the true selection function (or equivalently a scaling of the true mean galaxy density) $\hat{W}^{\text{resc}}(\bm x)= \mathcal{K}_{\text{survey}} / \mathcal{K} W(\bm x)$ where $\mathcal{K}_{\text{survey}} \equiv  N\left(\left. <R_{\mathcal{D}} \right|  \bm x =\bm  0 \right)   /  (\frac{4}{3} \pi R_{\mathcal{D}}^3) $ is the mean galaxy density within the survey region. 
The corresponding \say{dressed} number count is 
\begin{align} \label{eq:NrRdecomposedhom}
N^{\text{hom} }_{\hat{W}^\text{resc}}(\left. <r \right| \bm x ) &= \mathcal{K}_{\text{survey}} V(r) = \frac{\mathcal{K}_{\text{survey}}}{\mathcal{K}} N^{\text{hom}}( <r ) \, , 
\end{align}
where $V(r)$ is the Euclidean volume given by (\ref{eq:Numbercountexample3lim}). 
This number count clearly satisfies the integral constraint (\ref{eq:intconstraint}) on the scale of the survey, and in the simplest possible way, preserving the uniformity of the selection function. 
We now construct an estimated survey selection function such that the estimated radial selection function is determined from data using the integral constraint (\ref{eq:integralRho21}), but where the estimated angular selection function is homogeneous and equal to the true angular selection function of this toy model set-up.  
Following section \ref{sec:decoupling} and section \ref{sec:noinhomrad}, we thus have 
\begin{align} \label{eq:radialwindow}
\sqrt{g} \, \mathcal{K} \hat{W}^{\text{rad}}(\rad,\theta,\phi) &= \frac{1}{4 \pi} \bar{\rho}(\rad) \, \sin(\theta)   \, , 
\end{align}
where $\bar{\rho}(\rad)$ is given by (\ref{eq:densityexamplepolarr3}). The corresponding homogeneous number count yields
\begin{align} \label{eq:NRseprad}
\hspace*{-0.2cm} N^{\text{hom} }_{\hat{W}^\text{rad}}(\left. <r \right| \bm x ) &\equiv  \int_{0}^{R_{\mathcal{D}}} \dd \rad'  \int_{0}^{\pi} \dd \theta' \int_{-\pi}^{\pi} \dd \phi' \, \,  \sqrt{g} \, \mathcal{K} \hat{W}^{\text{rad}} (\rad',\theta',\phi')\, \Theta(r - d((\rad', \theta', \phi'),\bm x)) \, .
\end{align} 
{The function (\ref{eq:NRseprad}) represents the number count of galaxies after all structure in the angular direction has been smoothed over.}  
Finally, we consider the survey selection function estimation procedure of section \ref{sec:noinhom}, where the selection function is given by (\ref{eq:rhoRseparableConstrained}) and resembles both the radial and the angular distribution of the original survey
\begin{align} \label{eq:rhoRseparableConstrainedexample3d}
\sqrt{g} \, \mathcal{K} \hat{W}^{\text{sep}}(\rad,\theta,\phi) &= \frac{1}{N(R_{\mathcal{D}}) } \bar{\rho}(\rad) \, \bar{\rho}_{R_{\mathcal{D}}}(\theta, \phi)   \, , 
\end{align}
with $\bar{\rho}(\rad)$ and $\bar{\rho}(\theta,\phi)$ given by (\ref{eq:densityexamplepolarr3}) and (\ref{eq:densityexamplepolarphi3}) respectively. The estimated homogeneous numbercount from this ditribution is 
\begin{align} \label{eq:NRsep}
\hspace*{-0.2cm}  N^{\text{hom} }_{\hat{W}^\text{sep}}(\left. <r \right| \bm x ) &\equiv  \int_{0}^{R_{\mathcal{D}}} \dd \rad'  \int_{0}^{\pi} \dd \theta' \int_{-\pi}^{\pi} \dd \phi' \, \,  \sqrt{g} \, \mathcal{K} \hat{W}^{\text{sep}} (\rad',\theta',\phi')\, \Theta(r - d((\rad', \theta', \phi'),\bm x)) \, .
\end{align}
{This function represents the number count of galaxies after all joint structure between the angular and the radial directions has been removed from the original distribution.} 
The estimated normalised number count (\ref{eq:Nnormavselection}) corresponding to (\ref{eq:NrRdecomposedtrue}), (\ref{eq:NrRdecomposedhom}), (\ref{eq:NRseprad}), and (\ref{eq:NRsep}) respectively yield
\begin{align} \label{eq:Numbercountbare3d}
& \hat{\mathcal{N}}^{\text{bare}}(<r) \equiv 
\frac{N\left( <r  \right) }{  N^{\text{hom} }_{\hat{W}^\text{bare}} \left( <r  \right)  } \, , \qquad \hat{\mathcal{N}}^{\text{resc}}(<r) \equiv  
\frac{N\left( <r  \right) }{  N^{\text{hom} }_{\hat{W}^\text{resc}} \left( <r  \right)  } \, , \\
& \hat{\mathcal{N}}^{\text{rad}}(<r) \equiv  
\frac{N\left( <r  \right) }{N^{\text{hom} }_{\hat{W}^\text{rad}} \left( <r  \right)} \, ,  \qquad \hat{\mathcal{N}}^{\text{sep}}(<r) \equiv  
\frac{N\left( <r  \right) }{N^{\text{hom} }_{\hat{W}^\text{sep}}\left( <r  \right)} 
\end{align}
as defined for $r < R_{\mathcal{D}}$, and where the average number counts in spheres are given by (\ref{eq:Numbercountavexample3}) and (\ref{eq:NumbercountavRexample3}). 
{The function $\hat{\mathcal{N}}^{\text{bare}}(<r)$ represents the bare normalised number count (true normalised number count) of the toy model set-up, whereas $\hat{\mathcal{N}}^{\text{resc}}(<r)$, $\hat{\mathcal{N}}^{\text{rad}}(<r)$, and $\hat{\mathcal{N}}^{\text{sep}}(<r)$ represent different dressed normalised number counts where the estimated selection functions are not equal to the true selection function (which is known and equal to unity within the survey domain in this toy model setup). The rescaled normalised number count $\hat{\mathcal{N}}^{\text{resc}}(<r)$ represents the situation where the window function is correctly assumed to be uniform over the survey domain, but where the mean galaxy density of the survey is assumed to be the mean galaxy density of the Universe as a whole. 
The normalised count $\hat{\mathcal{N}}^{\text{rad}}(<r)$ represents the situation where the window function is correctly assumed to be uniform over the unit sphere, but where the structure of the galaxies contained in the survey in the radial direction is identified as a selection effect. Finally, the normalised count $\hat{\mathcal{N}}^{\text{sep}}(<r)$ represents the situation where structure of the galaxies in the purely radial \emph{and} in the purely angular direction is identified as a selection effect.}

\begin{figure}[!htb]
\centering
\begin{subfigure}[b]{.47\textwidth}
\addtocounter{subfigure}{0}
\includegraphics[width=\textwidth]{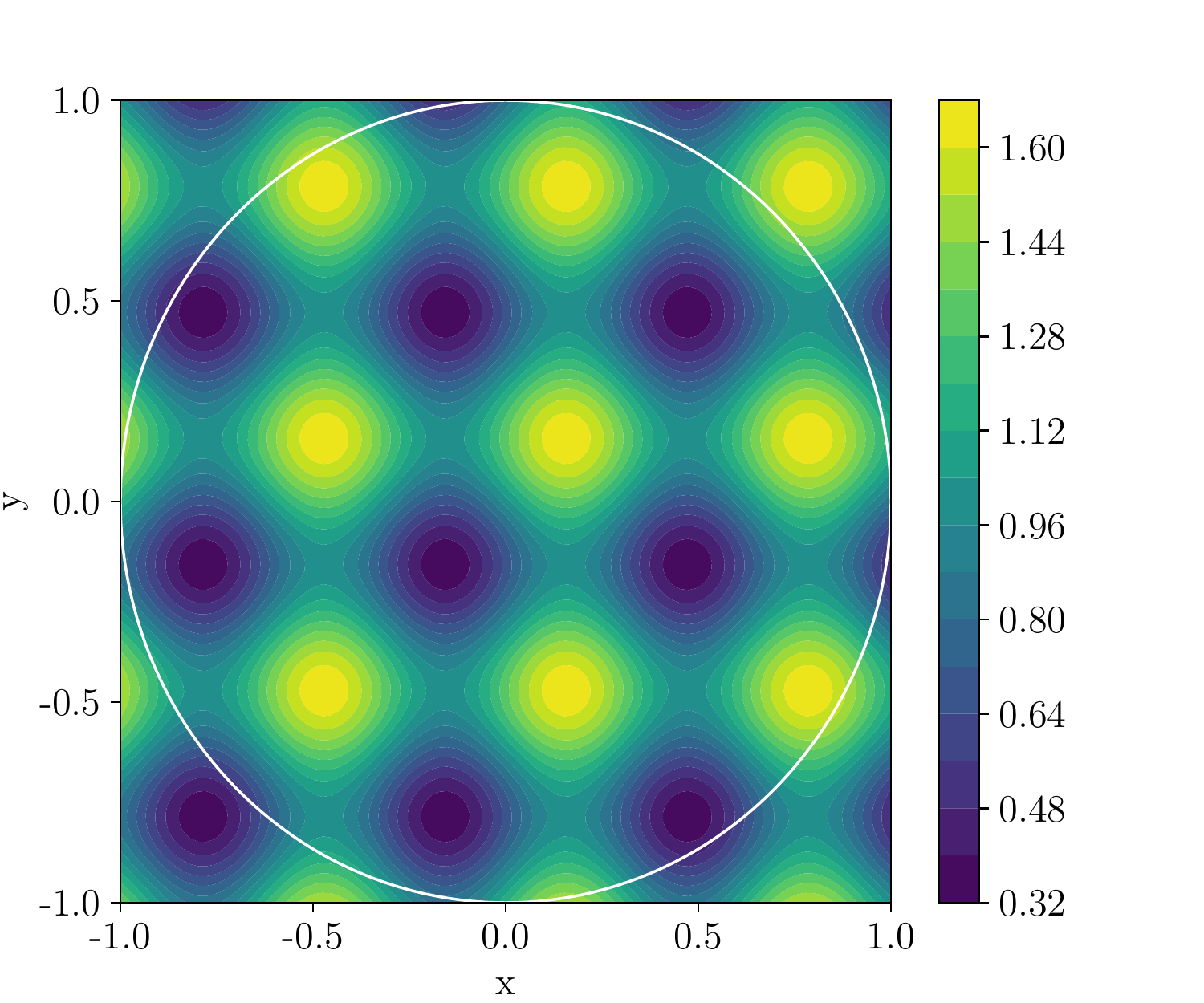}
\caption{Cross section $\rho(x,y) \, ;$ $\rho(\bm 0) = \mathcal{K}$} 
\label{subfig:density_3D_rhocenter1}
\end{subfigure}
\medskip
\begin{subfigure}[b]{.47\textwidth}
\addtocounter{subfigure}{2}
\includegraphics[width=\textwidth]{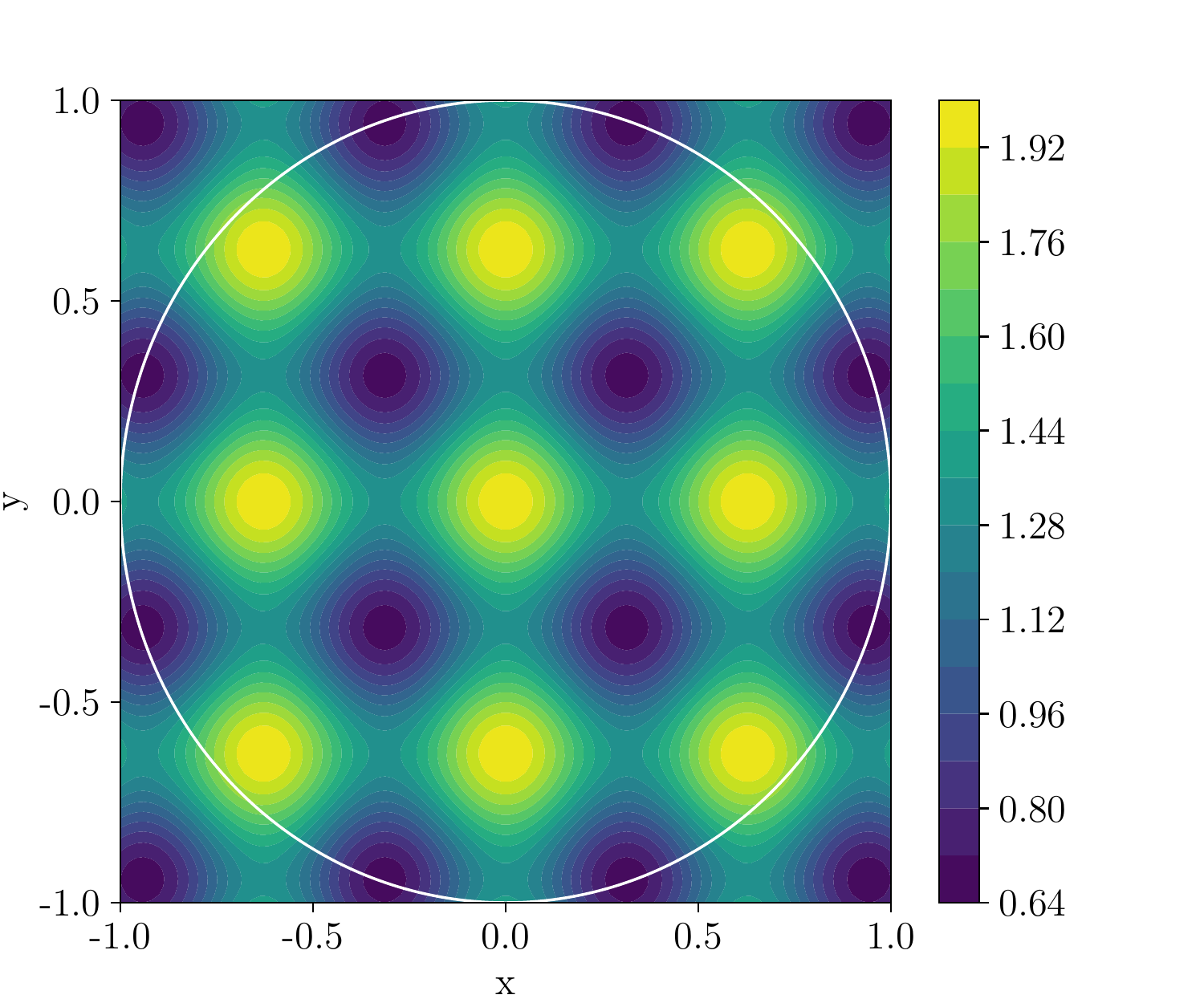}
\caption{Cross section $\rho(x,y) \, ;$ $\rho(\bm 0) = 2 \mathcal{K}$} 
\label{subfig:density_3D}
\end{subfigure}
\medskip
\begin{subfigure}[b]{.47\textwidth}
\addtocounter{subfigure}{-3}
\includegraphics[width=\textwidth]{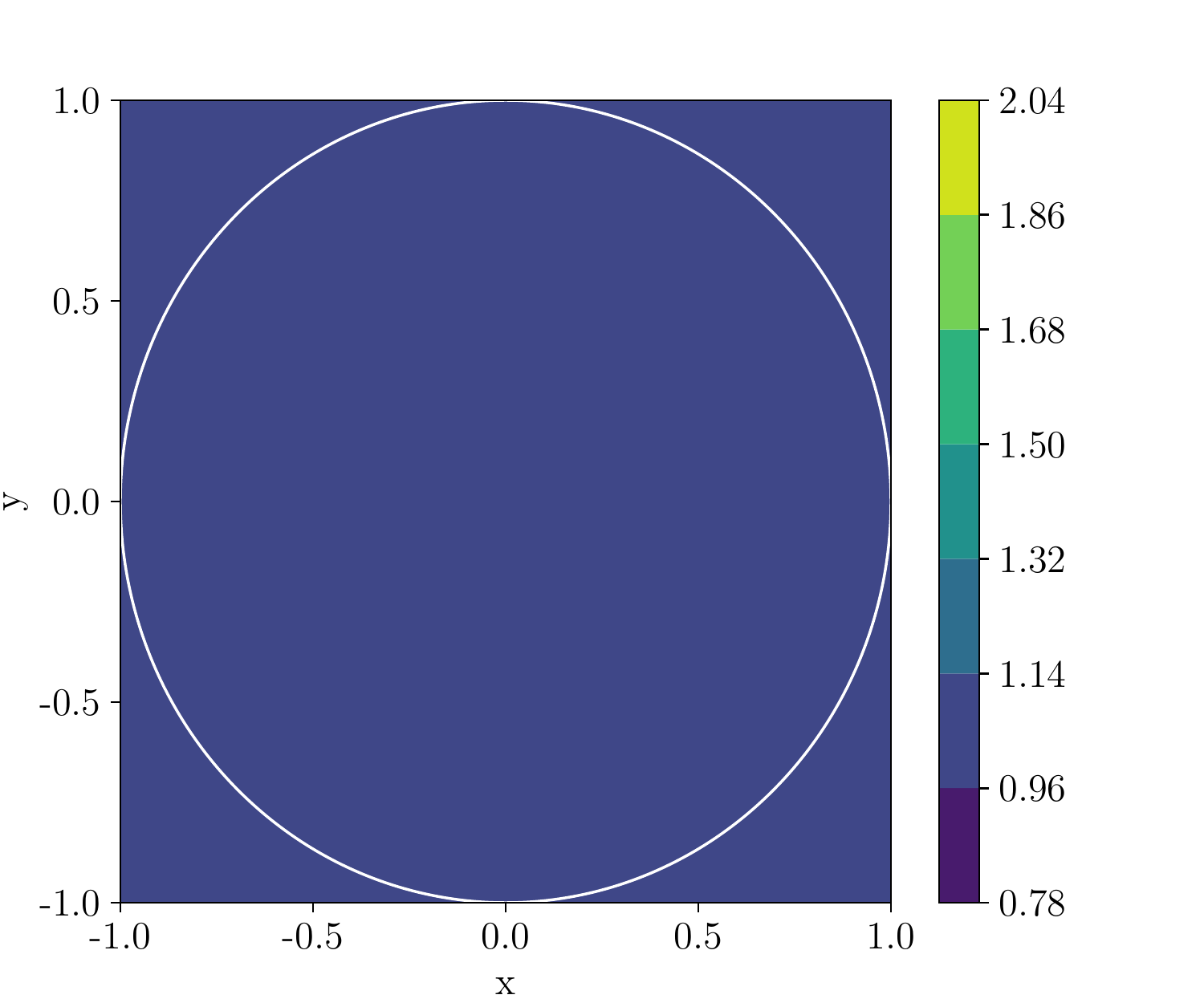}
\caption{Cross section $\mathcal{K} \hat{W}^{\text{rad}}(x,y) \, ;$ $\rho(\bm 0) = \mathcal{K}$}
\label{subfig:densityrad_3D_rhocenter1}
\end{subfigure}
\medskip
\begin{subfigure}[b]{.47\textwidth}
\addtocounter{subfigure}{2}
\includegraphics[width=\textwidth]{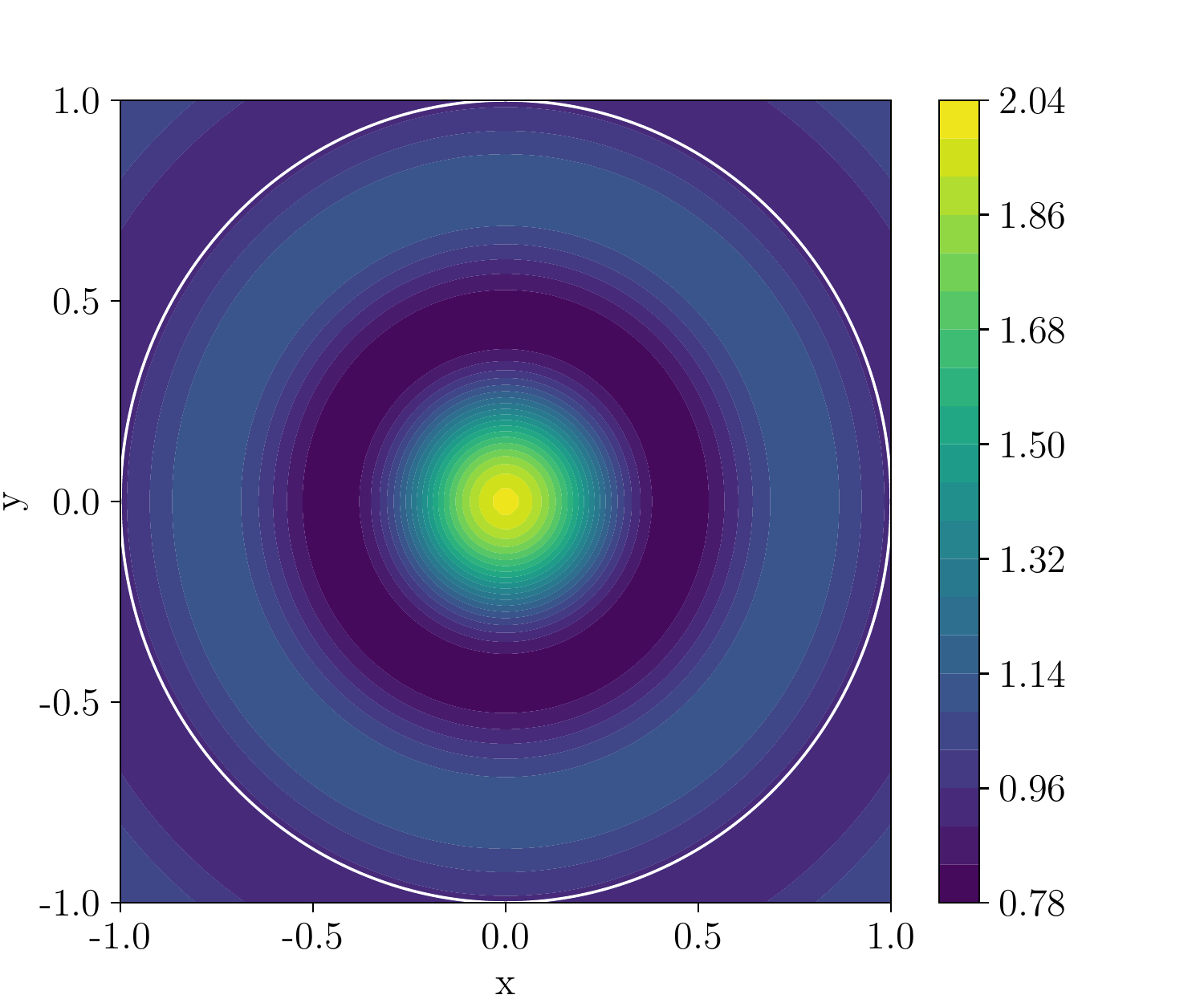}
\caption{Cross section $\mathcal{K} \hat{W}^{\text{rad}}(x,y) \, ;$ $\rho(\bm 0) = 2 \mathcal{K}$}
\label{subfig:densityrad_3D}
\end{subfigure}
\medskip
\begin{subfigure}[b]{.47\textwidth}
\addtocounter{subfigure}{-3}
\includegraphics[width=\textwidth]{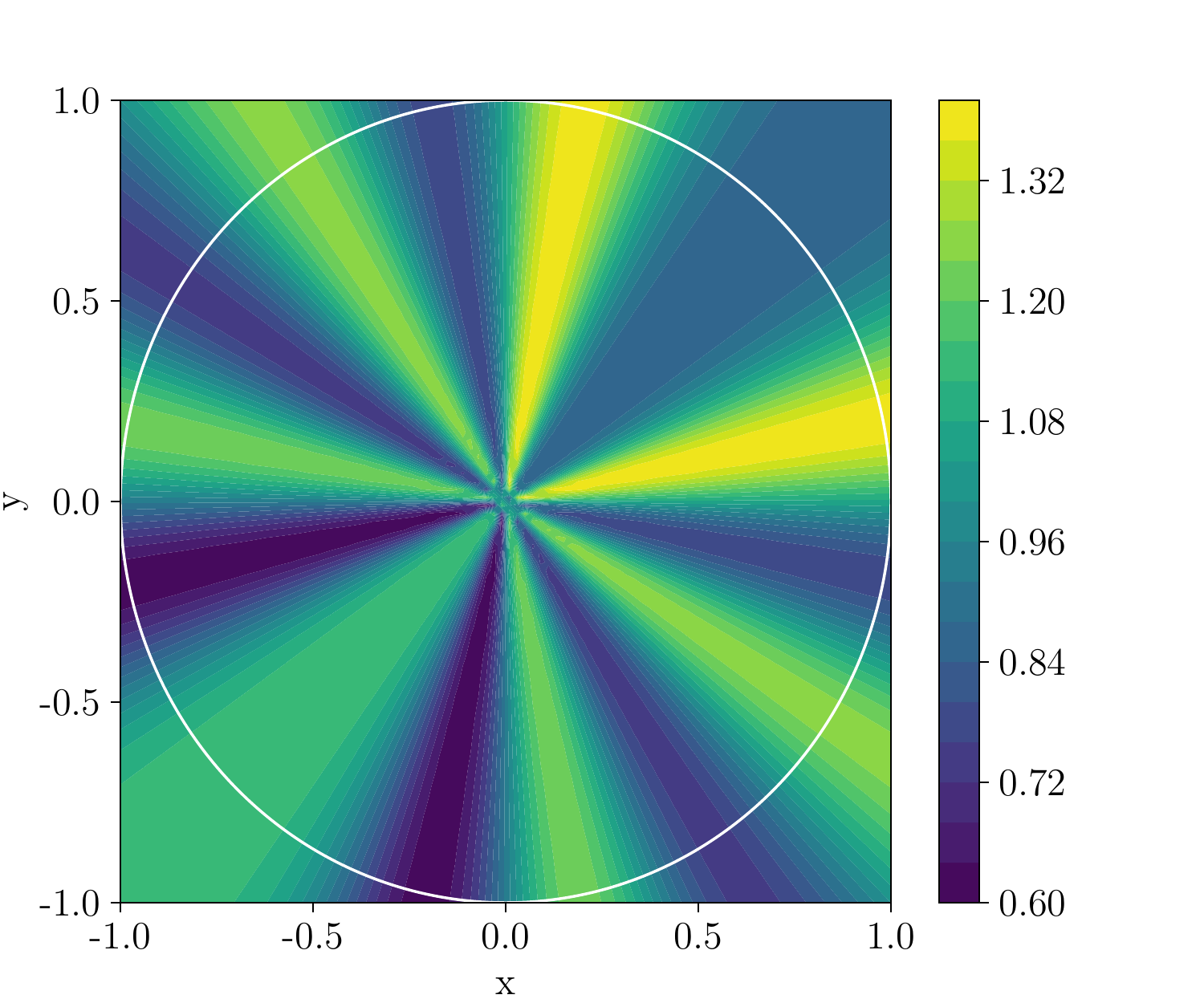}
\caption{Cross section $\mathcal{K} \hat{W}^{\text{sep}}(x,y) \, ;$ $\rho(\bm 0) = \mathcal{K}$}
\label{subfig:densitysep_3D_rhocenter1}
\end{subfigure}
\medskip
\begin{subfigure}[b]{.47\textwidth}
\addtocounter{subfigure}{2}
\includegraphics[width=\textwidth]{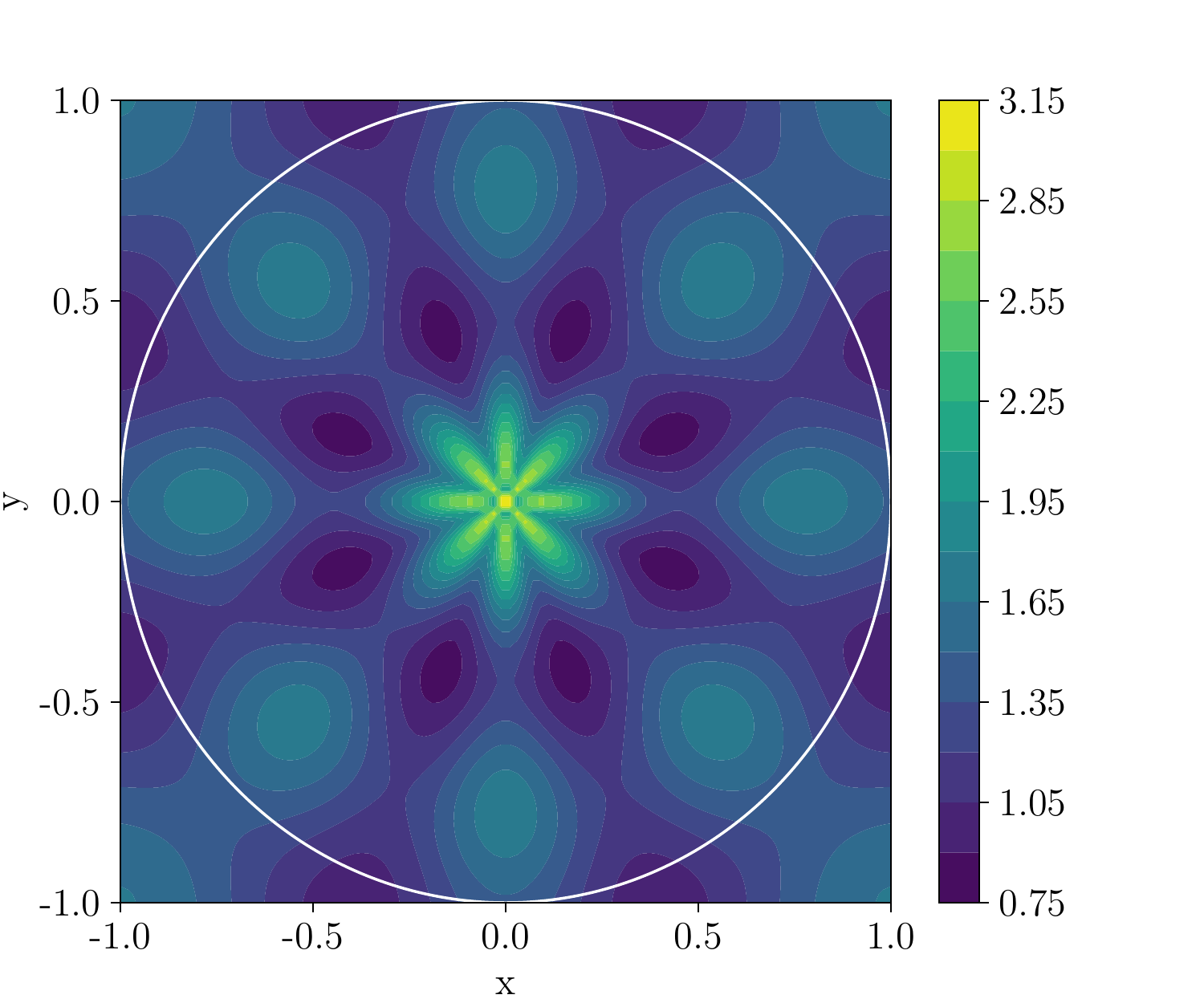}
\caption{Cross section $\mathcal{K} \hat{W}^{\text{sep}}(x,y) \, ;$ $\rho(\bm 0) = 2 \mathcal{K}$}
\label{subfig:densitysep_3D}
\end{subfigure}
\caption{Cross section $\rho(x,y)\equiv\rho(x,y,z=0)$ of the galaxy density and of the artificial galaxy densities $\mathcal{K} \hat{W}^{\text{rad}}(x,y)\equiv \mathcal{K} \hat{W}^{\text{rad}}(x,y,z=0)$ and $\mathcal{K} \hat{W}^{\text{sep}}(x,y)\equiv \mathcal{K} \hat{W}^{\text{sep}}(x,y,z=0)$ for a domain centered at a point of {average density (figures \ref{subfig:density_3D_rhocenter1}, \ref{subfig:densityrad_3D_rhocenter1}, \ref{subfig:densitysep_3D_rhocenter1}) and maximal density (figures \ref{subfig:density_3D}, \ref{subfig:densityrad_3D}, \ref{subfig:densitysep_3D}).} Parameters of $\rho(x,y,z)$ are $B=10$ and $\mathcal{K} = 1$. The radial domain size is $R_{\mathcal{D}} = 1$ and the domain boundary is shown as a circle on the plots.}
\label{fig:densities}
\end{figure}

Figure \ref{fig:densities} shows cross sections for the galaxy density and estimated selection functions for angular frequency $B=10$ and average galaxy density $\mathcal{K} = 1$. 
{In figure \ref{subfig:density_3D_rhocenter1} the cross section of the galaxy density field $\rho(x,y)\equiv\rho(x,y,z=0)$ is shown for a survey domain of radius $R_{\mathcal{D}} = 1$ centered on the point $\overset{o}{x} = \overset{o}{y} = \overset{o}{z} = 0$ of average density.  
Figure \ref{subfig:densityrad_3D_rhocenter1} shows the radial selection function (\ref{eq:radialwindow}) resembling the radial galaxy distribution of the domain, and figure \ref{subfig:densitysep_3D_rhocenter1} shows the multiplicative separable selection function (\ref{eq:rhoRseparableConstrainedexample3d}) resembling the radial \emph{and} the angular galaxy distribution.   
Figures \ref{subfig:density_3D}, \ref{subfig:densityrad_3D}, and \ref{subfig:densitysep_3D} show the same cross sections, but for a domain centered on a point of maximal density.  
}  
For a survey centered on a point of average galaxy density, the radial distribution $\bar{\rho}(\rad)$ is uniform, and only the angular part of $\hat{W}^{\text{sep}}(x,y)$ is non-trivial. 
For general survey centers both angular and radial selection account for structure in the physical matter distribution. As $B\rightarrow \infty$, while keeping the survey radius $R_{\mathcal{D}}$ constant, the inhomogeneous contributions in (\ref{eq:densityexamplepolarr3}) and (\ref{eq:densityexamplepolarphi3}) vanish, and the estimated selection functions (\ref{eq:radialwindow}) and (\ref{eq:rhoRseparableConstrainedexample3d}) reduce to the true homogeneous selection function {of this toy model setup: $\hat{W}^\text{rad} (\bm x) \rightarrow W(\bm x)$, $\hat{W}^\text{sep} (\bm x) \rightarrow W(\bm x)$. 
}  

\begin{figure}[!htb]
\centering
\begin{subfigure}[b]{.79\textwidth}
\includegraphics[width=\textwidth]{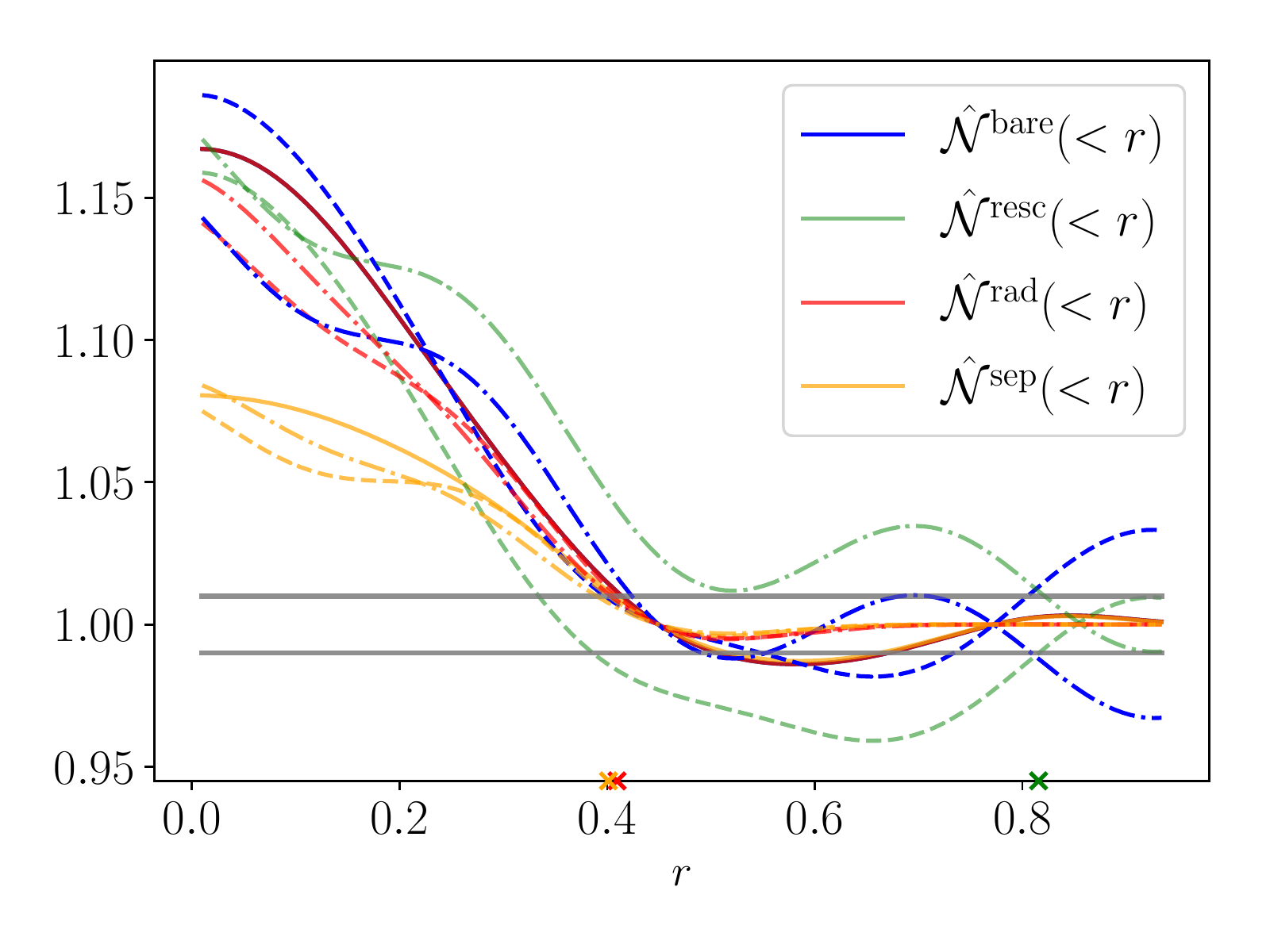}
\caption{Angular frequency $B=10$.}
\label{subfig:B10sphere}
\end{subfigure}
\medskip
\begin{subfigure}[b]{.79\textwidth}
\includegraphics[width=\textwidth]{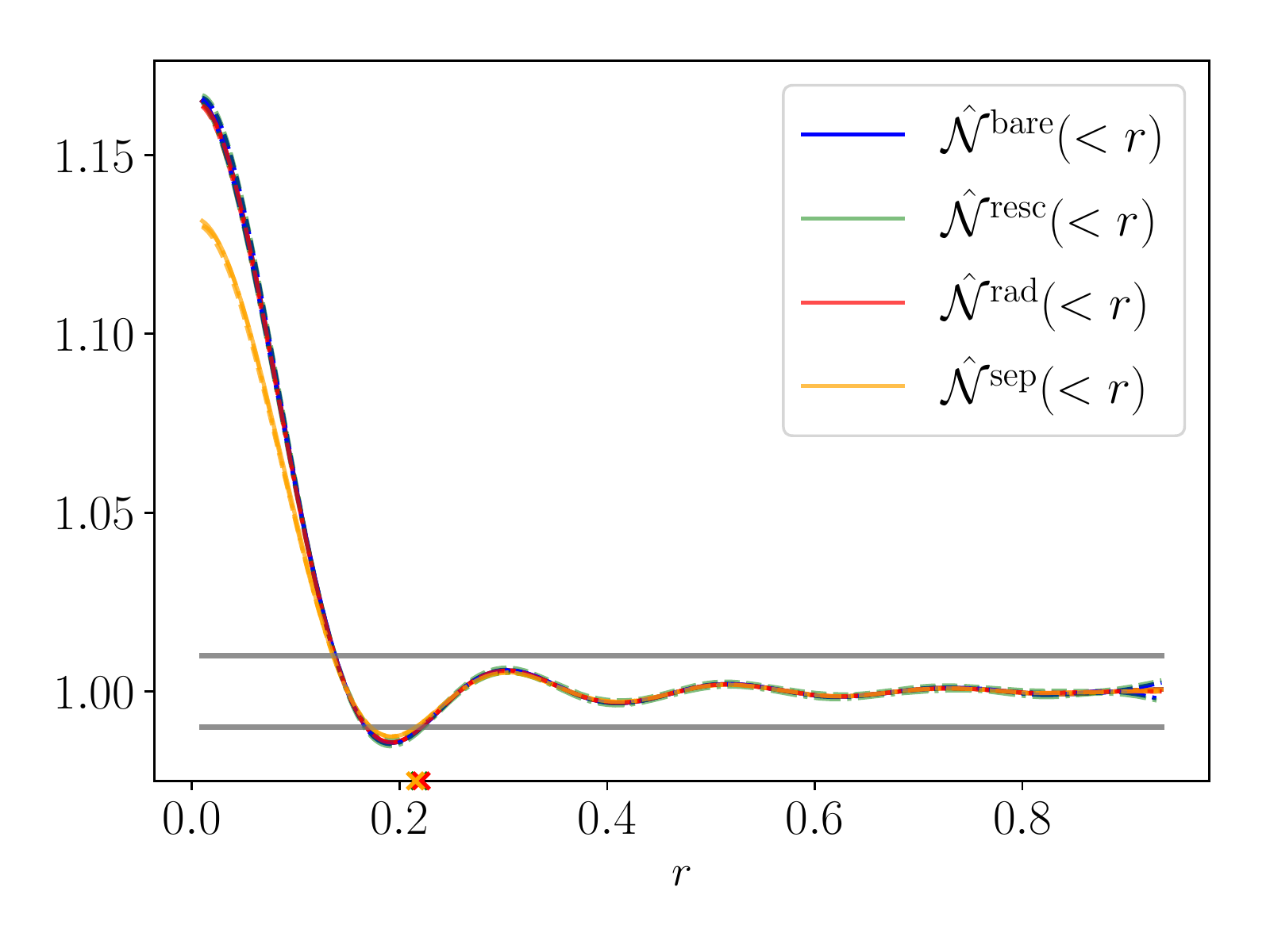}
\caption{Angular frequency $B=30$.}
\label{subfig:B30sphere}
\end{subfigure}
\caption{Scaled number counts $\hat{\mathcal{N}}^{\text{bare}}(<r)$, $\hat{\mathcal{N}}^{\text{resc}}(<r)$, $\hat{\mathcal{N}}^{\text{rad}}(<r)$, and $\hat{\mathcal{N}}^{\text{sep}}(<r)$ as a function of sphere radius $r$ for an spherical domain with radius $R_{\mathcal{D}} = 1$. Full drawn lines correspond to a survey centered at a point of average galaxy density $\mathcal{K}$. Dashed lines correspond a survey centered on a point of maximal galaxy density $2\mathcal{K}$, while dashed dotted lines correspond to a survey centered on a point of galaxy density equal to zero. {The grey horizontal lines show the homogeneity threshold $1 \pm 0.01$. The crosses on the $r$-axis illustrate the transition below 1\% for the dressed number counts for a survey centered on a point of maximal galaxy density.}
}
\label{fig:B103D}
\end{figure}

Figure \ref{fig:B103D} shows the bare number count and the dressed number counts for domain radius $R_{\mathcal{D}} = 1$ for three different choices of centers of the domain.  
Figure \ref{subfig:B10sphere} shows the number counts for an angular frequency of the oscillation $B=10$ {and figure \ref{subfig:B30sphere} shows the number counts for $B=30$. As is visible from the plots, the choice of frequency of the density oscillation is important for the level of agreement between the bare and dressed counts.  

We first discuss figure \ref{subfig:B10sphere}} with the period of the oscillation $T = \frac{2\pi}{B} = \frac{2\pi}{10}$ being smaller but comparable to the radius of the domain. {(For a discussion of the length scales in the toy model setup in relation to physical sizes of the largest structures and the volume coverage of typical surveys, see section \ref{sec:discussion}.)} 
From (\ref{eq:Numbercountexample3}) we have that the total number count within the domain can maximally differ $2.4\%$ from the homogeneous background number count $N^{\text{hom} }(<R_{\mathcal{D}})$, the exact difference depending on the galaxy density in the center of the survey. {Thus, the large scale homogeneity approximation (\ref{eq:assympthomD}) is violated with errors of $2.4\%$ or less on the scale of the survey. 
The violation of the cross sectional homogeneity approximations (\ref{eq:assympthomDranglestheory}), (\ref{eq:integralRho2theory}) are in general larger than of (\ref{eq:assympthomD}) with typical errors of order $20\%$.}
By construction the dressed number counts $\hat{\mathcal{N}}^{\text{resc}}(<r)$, $\hat{\mathcal{N}}^{\text{rad}}(<r)$, and $\hat{\mathcal{N}}^{\text{sep}}(<r)$ converge to $1$ for scales approaching the radius of the survey\footnote{{The convergence of $\hat{\mathcal{N}}^{\text{resc}}(<r)$ to $1$ is not fully visible in figure \ref{subfig:B10sphere}, which is due to the plot being truncated at $r\approx 0.93$ because of numerical singularities associated with the limit $r\rightarrow R_{\mathcal{D}}$.}}. Convergence is visibly not reached for the bare number counts for scales probed by the survey (due to a combination of the large periodicity of the density oscillation and the volume limitation of the survey), except for domain centers located at points close to mean galaxy density where convergence happens within the survey volume due the idealised spherical survey geometry investigated.
Because of the large errors associated with the approximations (\ref{eq:assympthomDranglestheory}) and (\ref{eq:integralRho2theory}) the selection functions $\hat{W}^{\text{rad}}(\bm x)$ and $\hat{W}^{\text{sep}}(\bm x)$ are associated with significant inhomogeneity and account for physical structure in the galaxy density field. As a result, the normalised number counts $\hat{\mathcal{N}}^{\text{rad}}(<r)$ and $\hat{\mathcal{N}}^{\text{sep}}(<r)$ are highly biased towards homogeneity as compared to the bare number count $\hat{\mathcal{N}}^{\text{bare}}(<r)$---the amplitude of $\hat{\mathcal{N}}^{\text{sep}}(<r)$ is especially suppressed with respect to the bare count. 
The oscillatory behaviour of the bare count is suppressed for the dressed counts $\hat{\mathcal{N}}^{\text{rad}}(<r)$ and $\hat{\mathcal{N}}^{\text{sep}}(<r)$, which exhibit smooth convergence.   
{The grey horizontal lines in figure \ref{subfig:B10sphere} show the 1\% inhomogeneity cutoff and the crosses on the $r$-axis mark the convergence of the dressed number counts below the 1\% threshold.}
Convergence below $1\%$ in amplitude of the normalised number counts happens at scales $r \sim 0.4$ for $\hat{\mathcal{N}}^{\text{rad}}(<r)$ and $\hat{\mathcal{N}}^{\text{sep}}(<r)$. For comparison, the true average number count in spheres $N(<r)$ converge below $1\%$ at scales $r = 0.664$ as detailed in section \ref{sec:exampleproperties}, yielding an error of approximately $40\%$ in the detected homogeneity scale.

Figure \ref{subfig:B30sphere} shows the same number counts as figure \ref{subfig:B10sphere}, but with the angular frequency of the oscillation increased to $B=30$. The period of the oscillation is in this case much smaller than the radius of the domain, and the mean density of the distribution is thus well sampled at the scale of the survey with maximal error of the approximation (\ref{eq:assympthomD}) of  $0.1\%$. The cross sectional homogeneity approximations (\ref{eq:assympthomDranglestheory}), (\ref{eq:integralRho2theory}) are associated with typical errors of $\sim 5\%$.
As a result, the dressed number counts coincide almost perfectly with the corresponding bare number counts, except for small scales $r\lesssim 0.1$ where the number counts $\hat{\mathcal{N}}^{\text{sep}}(<r)$ have slightly smaller amplitude. 
Convergence to of the normalised number counts below $1\%$ happens at the scale $r \sim 0.22$, which coincides with the true scale of convergence of $N(<r)$ below $1\%$ discussed in section \ref{sec:exampleproperties}.

The results in the present section are dependent on the spherical survey geometry. 
Replacing the geometry with that of an spherical shell of the same volume as the sphere, surfaces of constant radius become large when the radii of the concentric spheres are increased. Thus, we expect $\hat{\mathcal{N}}^{\text{rad}}(<r)$ to approach the bare number count $\hat{\mathcal{N}}^{\text{bare}}(<r)$ for an spherical shell of large radii, whereas we expect $\hat{\mathcal{N}}^{\text{sep}}(<r)$ to be strongly suppressed in amplitude due to the narrow radial width of the survey covered.  
Considering instead a cone-shaped survey of limited angular coverage---but extended in radius to preserve the volume of the sphere---we expect the amplitude of $\hat{\mathcal{N}}^{\text{rad}}(<r)$ to be more suppressed relative to $\hat{\mathcal{N}}^{\text{bare}}(<r)$ the more the angular coverage is decreased.

\subsection{Discussion of the implications of the results}
\label{sec:discussion} 
Let us interpret the results of section \ref{sec:spherical} with distances in units of $0.5\,$Gpc/h. The radius of the spherical survey in these units is $R_{\mathcal{D}} = 0.5\,$Gpc/h, and the angular frequencies investigated are $B=20 \,$h/Gpc and $B=60\,$h/Gpc. 
The survey has volume $4 \pi /24  (\text{Gpc/h})^3\sim 0.5 (\text{Gpc/h})^3$, which is comparable to the survey volumes in \cite{Hogg,Scrimgeour} used to probe convergence towards homogeneity. The total volume probed by the more recent SDSS-III BOSS CMASS and LOWZ surveys is $\sim  5 (\text{Gpc/h})^3$ \cite{Cuesta}, and is comprised of the CMASS North, CMASS South, LOWZ North, and LOWZ South surveys. These four subsamples have independently calibrated selection functions and must be treated separately when investigating systematic errors in selection function estimation.   

For the standard $\Lambda$CDM cosmology, a comoving radial scale $R_{\mathcal{D}} = 0.5\,$Gpc/h corresponds to redshifts of $z\sim0.2$, comparable to redshifts probed by the SDSS-III BOSS LOWZ survey \cite{Reid}. 
The period of the galaxy density oscillation is given as $T = 2\pi/B\;$, which for $B=20 \,$h/Gpc gives $T \sim 0.3 \;$Gpc/h. 
Based on the analysis in section \ref{sec:spherical} we thus expect structure of length scale comparable to or larger than $0.3 \;$Gpc/h to be poorly probed by normalised number count estimates $\hat{\mathcal{N}}^{\text{rad}}(<r)$, and $\hat{\mathcal{N}}^{\text{sep}}(<r)$, whereas scales $\ll 0.3 \;$Gpc/h are expected to be well-probed. 
For the BOSS LOWZ and CMASS surveys typical void sizes are of comoving diameter $\lesssim 160\,$Mpc/h  \cite{PMSutter,Mao}---depending on the exact criteria and void finding algorithm used. 
Based on this crude order of magnitude estimate, there is thus a possibility that the largest structures are poorly probed in number count estimators within the current largest surveys. 

{$\Lambda$CDM measurements of the BAO length scale are of the same order as current estimates of the scale of transition to 1\% inhomogeneity in average numbercounts. It might thus be suspected that normalisation of number count statistics could have implications for inference of the BAO feature in the galaxy distribution in addition to the homogeneity scale estimates investigated in this paper. One might investigate 2-point correlation function statistics to assess the potential impact on the BAO feature in a toy model setting similar to that of this paper.}

The conclusions on the sensitivity of the biasing of number counts from the imposed survey selection functions depend on the shape of the domain and on how the simplistic toy model investigations carry over to more realistic galaxy density distributions. 
In the toy model galaxy density distribution of section \ref{sec:exampleproperties}, the density contrast $(\rho - \mathcal{K})/\mathcal{K}$ was fixed to oscillate between $0$ and $1$ for simplicity. Decreasing (increasing) density contrasts, is expected to yield the same tendencies as found in the present toy model studies but with smaller (larger) deviations between the dressed and bare number counts.  

A general expectation from the investigations is that structure of size of order or larger than characteristic length scales of the survey are not accurately probed when invoking typical angular and radial survey selection functions to normalise number counts. 
This can be the case even if the survey covers a volume within which the mean density of the distribution is well sampled, if the survey is sufficiently limited in either radial or angular coverage. 
For instance, surveys of small angular coverage could potentially constitute an issue for homogeneity scale estimates based on the calibrated number count $\hat{\mathcal{N}}^{\text{rad}}(<r)$, since angular cross sections of the survey might not sample the average galaxy density well resulting in break down of the approximation (\ref{eq:integralRho2theory1}). 
Similarly for surveys that have a depth in redshift of order a few times that of the largest physical structures or smaller, might have significant physical inhomogeneity in the projected angular distribution function and thus yield dressed number counts $\hat{\mathcal{N}}^{\text{sep}}(<r)$ of considerable error. 
We note in this regard that the samples used to probe homogeneity in \cite{Hogg} and \cite{Scrimgeour} are of angular coverage $\sim 4,000 \, (\text{deg})^2$ and $\sim 1,000 \, (\text{deg})^2$ respectively, corresponding to $\sim 6\%$ and $\sim 2\%$ of the total sky. 
For upcoming data from the DESI experiment \cite{DESI}, the angular coverage will be $\sim 14,000 \, (\text{deg})^2$ and with a redshift coverage for luminous red galaxies of $z \lesssim 1$. 
The bigger the coverage of the survey (and the smaller the physical size of the largest structures in the Universe), the less sensitive we expect homogeneity scale estimates to be on the exact approximation of the selection function used.

\section{Conclusion}
\label{sec:conclusion}
We have investigated number counts in spheres as probes of transition to homogeneity in the matter distribution.
Normalised number counts---where the normalisation is computed from an assumed survey selection function---are used in modern analysis to account for selection effects. 
Typical selection functions employed for large scale galaxy surveys assume convergence to homogeneity on the largest scales of the survey. 
When such estimated selection functions are used for constructing number count in spheres statistics, there is a risk of underestimating the level of inhomogeneity at a given scale. 
While investigations using $\Lambda$CDM {simulations} to determine the impact of the selection function estimation \cite{Burden,Mattia} are relevant for universe models where the largest structures are well described by the $\Lambda$CDM prediction of structure formation, such studies should not be used to derive model independent statements or when the aim is to consistency test the $\Lambda$CDM paradigm. 

The modelling of the survey selection function constitutes an additional source of bias of homogeneity scale estimates to that of (i) the incompleteness of spheres and the use of artificial catalogoues to account for the incompleteness \cite{Scrimgeour,Laurent} (ii) tracers as biased probes of the mass distribution \cite{Desjacques,White} (iii) galaxy evolution \cite{Peng} and (iv) finite sampling and finite resolution in cosmological datasets \cite{Poissonfractal,Boxcounting}. 

We have considered an example of an oscillating galaxy density distribution with a notion of large scale convergence to spatial homogeneity, and considered bare number counts (true number counts) as well as dressed number counts (number counts corrected for structure in the radial and angular direction). 
For periods of the oscillation much smaller than the radial and angular scales of the survey---with convergence to homogeneity approximately satisfied at the largest scales of the survey---the bare number counts were well approximated by the dressed number counts.    
However, for periods of the oscillation comparable to scales of the survey, typical correction procedures for selection effects where shown to artificially yield number counts dressed towards homogeneity.
This indicate that the largest structures in our Universe might not be well-probed by conventional homogeneity scale estimators, if these are employed within survey domains of radial or angular coverage of similar order of magnitude as the physical size of the structures. 
Without independent determination of the survey selection function, we might think of the resulting homogeneity scale estimates as providing lower bounds for the transition scale to the given level of homogeneity.

\acknowledgments
This work is part of a project that has received funding from the European Research Council (ERC) under the European Union's Horizon 2020 research and innovation programme (grant agreement ERC advanced grant 740021--ARTHUS, PI: Thomas Buchert). I wish to thank Chris Blake, Thomas Buchert, Martin Kerscher and Pierre Mourier for useful comments. 
{I thank the referee for constructive comments, which contributed to improving this paper.}


\begin{thebibliography}{99}

\bibitem{Cole}
Cole~S \etal,
{The 2dF Galaxy Redshift Survey: Power-spectrum analysis of the final dataset and cosmological implications},
\href{https://doi.org/10.1111/j.1365-2966.2005.09318.x}{\MNRAS{362} (2005) 505}
[\link{astro-ph/0501174}]

\bibitem{EisensteinDetection}
Eisenstein~D~J \etal,
{Detection of the Baryon Acoustic Peak in the Large-Scale Correlation Function of SDSS Luminous Red Galaxies},
\href{https://doi.org/10.1086/466512}{\ApJ{633} (2005) 560}
[\link{astro-ph/0501171}]

\bibitem{Hogg}
Hogg~D et al.,
Cosmic homogeneity demonstrated with luminous red galaxies
\href{https://doi.org/10.1086/429084}{\ApJ{624} (2005) 54}   
[\href{https://arxiv.org/abs/astro-ph/0411197}{arXiv:astro-ph/0411197}] 

\bibitem{Scrimgeour}
Scrimgeour~M et al.,
The WiggleZ Dark Energy Survey: the transition to large-scale cosmic homogeneity
\href{https://doi.org/10.1111/j.1365-2966.2012.21402.x}{\MNRAS{425} (2012) 116}   
[\href{https://arxiv.org/abs/1205.6812}{arXiv:1205.6812}] 

\bibitem{Laurent}
Laurent~P et al.,
A 14 $h^{-3}$ Gpc${}^3$ study of cosmic homogeneity using BOSS DR12 quasar sample
\href{https://doi.org/10.1088/1475-7516/2016/11/060}{\JCAP{11} (2016) 060}   
[\href{https://arxiv.org/abs/1602.09010}{arXiv:1602.09010}] 

\bibitem{BlakeBrough}
Blake~C et al.,
The WiggleZ Dark Energy Survey: the selection function and z=0.6 galaxy power spectrum
\href{https://doi.org/10.1111/j.1365-2966.2010.16747.x}{\MNRAS{406} (2010) 803}   
[\href{https://arxiv.org/abs/1003.5721}{arXiv:1003.5721}] 

\bibitem{Reid}
Reid~B et al.,
SDSS-III Baryon Oscillation Spectroscopic Survey Data Release 12: galaxy target selection and large scale structure catalogues
\href{https://doi.org/10.1093/mnras/stv2382}{\MNRAS{455} (2016) 1553}   
[\href{https://arxiv.org/abs/1509.06529}{arXiv:1509.06529}] 

\bibitem{Eisensteintarget}
Eisenstein~D~J \etal,
{Spectroscopic target selection for the Sloan Digital Sky Survey: The Luminous red galaxy sample},
\href{https://doi.org/10.1086/323717}{\AJ{122} (2001) 2267}   
[\href{https://arxiv.org/abs/astro-ph/0108153}{astro-ph/0108153}]

\bibitem{Zehavi}
Zehavi~I et al.,
The Intermediate-scale clustering of luminous red galaxies
\href{https://doi.org/10.1086/427495}{\ApJ{621} (2005) 22}   
[\href{https://arxiv.org/abs/astro-ph/0411557}{astro-ph/0411557}]

\bibitem{Ata}
Ata~M et al.,
The clustering of the SDSS-IV extended Baryon Oscillation Spectroscopic Survey DR14 quasar sample: first measurement of baryon acoustic oscillations between redshift 0.8 and 2.2
\href{https://doi.org/10.1093/mnras/stx2630}{\MNRAS{473} (2018) 4773}   
[\href{https://arxiv.org/abs/1705.06373}{arXiv:1705.06373}] 

\bibitem{DESI}
Aghamousa~A et al., 
DESI Experiment Part I: Science,Targeting, and Survey Design
\emph{arXiv e-prints (2016)}
[\href{https://arxiv.org/abs/1611.00036}{arXiv:1611.00036}] 

\bibitem{Burden}
Burden~A et al., 
Mitigating the Impact of the DESI Fiber Assignment on Galaxy Clustering 
\href{https://doi.org/10.1088/1475-7516/2017/03/001}{\JCAP{03} (2017) 001}   
[\href{https://arxiv.org/abs/1611.04635}{arXiv:1611.04635}] 

\bibitem{Goncalves}
Gon\c calves~R~S, Carvalho~G~C, Bengaly~C~A~P, Carvalho~J~C and Alcaniz~J~S, 
Measuring the scale of cosmic homogeneity with SDSS-IV DR14 quasars
\href{https://doi.org/10.1093/mnras/sty2670}{\MNRAS{481} (2018) 4}   
[\href{https://arxiv.org/abs/1809.11125}{arXiv:1809.11125}] 

\bibitem{Ntelis}
Ntelis~P et al.,
The scale of cosmic homogeneity as a standard ruler
\href{10.1088/1475-7516/2018/12/014}{\JCAP{12} (2018) 014}   
[\href{https://arxiv.org/abs/1810.09362}{arXiv:1810.09362}] 

\bibitem{Avila1}
Avila~F, Novaes~C~P, Bernui~A and Carvalho~E~de,
The scale of homogeneity in the local Universe with the ALFALFA catalogue
\href{https://doi.org/10.1088/1475-7516/2018/12/041}{\JCAP{12} (2018) 041}   
[\href{https://arxiv.org/abs/1806.04541}{arXiv:1806.04541}] 

\bibitem{Avila2}
Avila~F, Novaes~C~P, Bernui~A, Carvalho~E~de and Nogueira-Cavalcante~J~P,
The angular scale of homogeneity in the Local Universe with the SDSS blue galaxies
\href{https://doi.org/10.1093/mnras/stz1765}{\MNRAS{488} (2019) 1481}   
[\href{https://arxiv.org/abs/1906.10744}{arXiv:1906.10744}] 

\bibitem{PeeblesNpoint}
Peebles~P~J~E,
The galaxy and mass n-point correlation functions: a blast from the past, 
{\em ASP Conf. Ser.} \textbf{252} (2001) 201
[\href{https://arxiv.org/abs/astro-ph/0103040}{astro-ph/0103040}] 

\bibitem{MeckeBuchertWagner}
Mecke~K~R, Buchert~T and Wagner~H,
Robust Morphological Measures for Large-Scale Structure in the Universe, 
{\em Astron. Astrophys.} \textbf{288} (1994) 697
[\href{https://arxiv.org/abs/astro-ph/9312028}{astro-ph/9312028}] 

\bibitem{Wiegand}
Wiegand~A, Buchert~T and Ostermann~M, 
Direct Minkowski Functional analysis of large redshift surveys: a new high--speed code tested on the luminous red galaxy Sloan Digital Sky Survey-DR7 catalogue
\href{https://doi.org/10.1093/mnras/stu1118}{\MNRAS{443} (2014) 241}   
[\href{https://arxiv.org/abs/1311.3661}{arXiv:1311.3661}] 

\bibitem{Hosoya}
Hosoya~A, Buchert~T and Morita~M, 
Information Entropy in Cosmology
\href{https://doi.org/10.1103/PhysRevLett.92.141302}{\PRL{92} (2004) 141302}   
[\href{https://arxiv.org/abs/gr-qc/0402076}{gr-qc/0402076}] 

\bibitem{Pandey}
Pandey~B, 
A method for testing the cosmic homogeneity with Shannon entropy
\href{https://doi.org/10.1093/mnras/stt134}{\MNRAS{430} (2013) 3376}   
[\href{https://arxiv.org/abs/1301.4961}{arXiv:1301.4961}] 

\bibitem{SarkarPandeygalaxies}
Pandey~B and Sarkar~S, 
Probing large scale homogeneity and periodicity in the LRG distribution using Shannon entropy
\href{https://doi.org/10.1093/mnras/stw1075}{\MNRAS{460} (2016) 1519}   
[\href{https://arxiv.org/abs/1512.06350}{arXiv:1512.06350}] 

\bibitem{SarkarPandeyquasars}
Sarkar~S and Pandey~B, 
An information theory based search for homogeneity on the largest accessible scale
\href{https://doi.org/10.1093/mnrasl/slw145}{\MNRAS{463} (2016) L12}   
[\href{https://arxiv.org/abs/1607.06194}{arXiv:1607.06194}] 

\bibitem{Beisbart}
Beisbart~C, Kerscher~M and Mecke~K, 
Mark Correlations: Relating Physical Properties to Spatial Distributions 
\href{https://ui.adsabs.harvard.edu/abs/2002LNP...600..358B/abstract}{Lecture Notes in Physics, 600, Springer, eds. K. Mecke \& D. Stoyan (2002)}   
[\href{https://arxiv.org/abs/physics/0201069}{physics/0201069}] 

\bibitem{ShethDiaferio}
Sheth~R~K and Diaferio~A, 
How unusual are the Shapley Supercluster and the Sloan Great Wall?
\href{https://doi.org/10.1111/j.1365-2966.2011.19453.x}{\MNRAS{417} (2011) 2938}   
[\href{https://arxiv.org/abs/1105.3378}{arXiv:1105.3378}] 

\bibitem{ParkSGW}
Park~C et al.,
The Challenge of the Largest Structures in the Universe to Cosmology 
\href{https://doi.org/10.1088/2041-8205/759/1/L7}{{\em Astrophys.\ J. Lett.} \textbf{759} (2012) 6}   
[\href{https://arxiv.org/abs/1209.5659}{arXiv:1209.5659}] 

\bibitem{Keenan}
Keenan~R~C,  Barger~A~J and Cowie~L~L,
Evidence for a $\sim$300 Megaparsec Scale Under-density in the Local Galaxy Distribution 
\href{https://doi.org/10.1088/0004-637X/775/1/62}{{\em Astrophys.\ J.} \textbf{775} (2013) 16}   
[\href{https://arxiv.org/abs/1304.2884}{arXiv:1304.2884}] 

\bibitem{Clowes}
Clowes~C~G et al., 
A structure in the early universe at z ~ 1.3 that exceeds the homogeneity scale of the R-W concordance cosmology
\href{https://doi.org/10.1093/mnras/sts497}{\MNRAS{429} (2013) 2910}   
[\href{https://arxiv.org/abs/1211.6256}{arXiv:1211.6256}] 

\bibitem{Park}
Park~C, Song~H, Einasto~M, Lietzen~H and Heinamaki~P,
Large SDSS quasar groups and their statistical significance
\href{https://doi.org/10.5303/JKAS.2015.48.1.075}{{\em J. Korean Astron. Soc.} \textbf{48} (2015) 75}   
[\href{https://arxiv.org/abs/1502.03563}{arXiv:1502.03563}] 

\bibitem{Marinello}
Marinello~G~E et al.,
Compatibility of the Large Quasar Groups with the Concordance Cosmological Model
\href{https://doi.org/10.1093/mnras/stw1513}{\MNRAS{461} (2016) 2267}   
[\href{https://arxiv.org/abs/1603.03260}{arXiv:1603.03260}] 

\bibitem{BolejkoOstrowski}
Bolejko~K, Ostrowski~J~J,
The environment-dependence of the growth of the most massive objects in the Universe
\href{https://doi.org/10.1103/PhysRevD.99.124036}{{\em Phys. Rev. D} (2019) 124036}   
[\href{https://arxiv.org/abs/1805.11047}{arXiv:1805.11047}] 

\bibitem{Feindt}
Feindt~U et al.,
Measuring cosmic bulk flows with Type Ia Supernovae from the Nearby Supernova Factory
\href{https://doi.org/10.1051/0004-6361/201321880}{\AaA{560} (2013) A90}   
[\href{https://arxiv.org/abs/1310.4184}{arXiv:1310.4184}] 

\bibitem{Magoulas}
Magoulas~C et al.,
Measuring the cosmic bulk flow with 6dFGSv 
\href{https://doi.org/10.1017/S1743921316010115}{{\em Proc. IAU Symp.} \textbf{308} (2016) 336}   

\bibitem{Colin}
Colin~J, Mohayaee~R, Sarkar~S and Shafieloo~A,
Probing the anisotropic local universe and beyond with SNe Ia data
\href{https://doi.org/10.1111/j.1365-2966.2011.18402.x}{\MNRAS{414} (2011) 264}   
[\href{https://arxiv.org/abs/1011.6292}{arXiv:1011.6292}] 

\bibitem{Labini}
Labini~F~S, 
Inhomogeneities in the universe
\href{https://doi.org/10.1088/0264-9381/28/16/164003}{\CQG{28} (2011) 164003}   
[\href{https://arxiv.org/abs/1103.5974}{arXiv:1103.5974}] 

\bibitem{LabiniVasilyevBaryshev}
Labini~S~L, Vasilyev~N~L, and Baryshev~Y~V, 
Breaking the self-averaging properties of spatial galaxy fluctuations in the Sloan Digital Sky Survey - Data Release Six
\href{https://doi.org/10.1051/0004-6361/200811565}{\emph{Astron. Astrophys.} \textbf{508} 17} 
[\href{https://arxiv.org/abs/0909.0132}{arXiv:0909.0132}] 

\bibitem{CabreGaztanaga}
Cabre~A and Gaztanaga~E, 
Clustering of luminous red galaxies I: large scale redshift space distortions
\href{https://doi.org/10.1111/j.1365-2966.2008.14281.x}{\MNRAS{393} (2009) 1183}   
[\href{https://arxiv.org/abs/0807.2460}{arXiv:0807.2460}] 

\bibitem{Mattia}
de Mattia~A and Ruhlmann-Kleider~V,
Integral constraints in spectroscopic surveys
\href{https://doi.org/10.1088/1475-7516/2019/08/036}{\JCAP{08} (2019) 036}   
[\href{https://arxiv.org/abs/1904.08851}{arXiv:1904.08851}] 

\bibitem{PeeblesTheory}
Peebles~P~J~E,
Statistical Analysis of Catalogs of Extragalactic Objects. I. Theory 
\href{https://doi.org/10.1086/152431}{\ApJ{185} (1973) 413}

\bibitem{Desjacques}
Desjacques~V et al.,
Large-Scale Galaxy Bias 
\href{https://doi.org/10.1016/j.physrep.2017.12.002}{\emph{Phys. Rept.} {\bf 733} (2018) 1}   
[\href{https://arxiv.org/abs/1611.09787}{arXiv:1611.09787}] 

\bibitem{White}
White~M et al.,
The clustering of intermediate-redshift quasars as measured by the Baryon Oscillation Spectroscopic Survey 
\href{https://doi.org/10.1111/j.1365-2966.2012.21251.x}{\MNRAS{424} (2012) 933}   
[\href{https://arxiv.org/abs/1203.5306}{arXiv:1203.5306}] 

\bibitem{Peng}
Peng~Y et al.,
Mass and Environment as Drivers of Galaxy Evolution in SDSS and zCOSMOS and the Origin of the Schechter Function
\href{https://doi.org/10.1088/0004-637X/721/1/193}{\ApJ{721} (2010) 193}   
[\href{https://arxiv.org/abs/1003.4747}{arXiv:1003.4747}] 

\bibitem{Poissonfractal}
Dubrulle~B and Lachieze-Rey~M,
The Poisson distribution on a fractal
\href{https://ui.adsabs.harvard.edu/abs/1994A\%26A...287..361D/abstract}{\AaA{287} (1994) 361}   

\bibitem{Boxcounting}
Dubrulle~B and Lachieze-Rey~M,
On the multifractal analysis of galaxy catalogs with box-counting methods
\href{https://ui.adsabs.harvard.edu/abs/1994A\%26A...289..667D/abstract}{\AaA{289} (1994) 667}   

\bibitem{Yadav}
Yadav~J et al.,
Testing homogeneity on large scales in the Sloan Digital Sky Survey Data Release One 
\href{https://doi.org/10.1111/j.1365-2966.2005.09578.x}{\MNRAS{364} (2005) 601}   
[\href{https://arxiv.org/abs/astro-ph/0504315}{astro-ph/0504315}] 

\bibitem{DPestimator}
Davis~M and Peebles~P~J~E,
A survey of galaxy redshifts. V. The two-point position and velocity correlations
\href{https://doi.org/10.1086/160884 }{\ApJ{267} (1983) 465}   

\bibitem{Kerscher}
Kerscher~M, Szapudi~I and Szalay~A, 
A comparison of estimators for the two-point correlation function
\href{https://doi.org/10.1086/312702}{\ApJ{535} (2000) L13}   
[\href{https://arxiv.org/abs/astro-ph/9912088}{astro-ph/9912088}]  

\bibitem{Myers:2015hpw}
Myers~A~D et al., 
The SDSS-IV extended Baryon Oscillation Spectroscopic Survey: Quasar Target Selection 
\href{https://doi.org/10.1088/0067-0049/221/2/27}{\emph{Astrophys. J. Suppl.} \textbf{221} (2015) 27}   
[\href{https://arxiv.org/abs/1508.04472}{arXiv:1508.04472}] 

\bibitem{Ross}
Ross~A~J et al.,
The clustering of galaxies in the SDSS-III Baryon Oscillation Spectroscopic Survey: analysis of potential systematics
\href{https://doi.org/10.1111/j.1365-2966.2012.21235.x}{\MNRAS{424} (2012) 564}   
[\href{https://arxiv.org/abs/1203.6499}{arXiv:1203.6499}] 

\bibitem{Guo}
Guo~H, Zehavi~I and Zheng~Z, 
A New Method to Correct for Fiber Collisions in Galaxy Two-Point Statistics  
\href{https://doi.org/10.1088/0004-637X/756/2/127}{\ApJ{756} (2012) 127}   
[\href{https://arxiv.org/abs/1111.6598}{arXiv:1111.6598}] 

\bibitem{SikoraGlod}
Sikora~S and G\l\'od~K 
Example of an inhomogeneous cosmological model in the context of backreaction 
\href{https://doi.org/10.1103/PhysRevD.95.063517}{\emph{Phys. Rev. D} \textbf{95} (2017) 063517} 
[\href{https://arxiv.org/abs/1612.03604}{arXiv:1612.03604}] 

\bibitem{Cuesta}
Cuesta~A~J et al.,
The clustering of galaxies in the SDSS-III Baryon Oscillation Spectroscopic Survey: Baryon Acoustic Oscillations in the correlation function of LOWZ and CMASS galaxies in Data Release 12
\href{https://doi.org/10.1093/mnras/stw066}{\MNRAS{457} (2016) 1770}   
[\href{https://arxiv.org/abs/1509.06371}{arXiv:1509.06371}] 

\bibitem{PMSutter}
Sutter~P~M, 
Voids in the SDSS DR9: observations, simulations, and the impact of the survey mask
\href{https://doi.org/10.1093/mnras/stu1094}{\MNRAS{442} (2014) 3127}   
[\href{https://arxiv.org/abs/1310.7155}{arXiv:1310.7155}] 

\bibitem{Mao}
Mao~Q, 
A Cosmic Void Catalog of SDSS DR12 BOSS Galaxies 
\href{https://doi.org/10.3847/1538-4357/835/2/161}{\ApJ{835} (2017) 161}   
[\href{https://arxiv.org/abs/1602.02771}{arXiv:1602.02771}] 



\end{thebibliography}
\end{document}